\documentstyle{mn}

\newif\ifAMStwofonts

\title{Hydrogen model atmospheres for white dwarf stars}

\author[R.D. Rohrmann]
{R. D. Rohrmann\thanks{Fellow of the Consejo Nacional de
Investigaciones Cient\'{\i}ficas y T\'ecnicas (CONICET), Argentina.
Email: rohr@oac.uncor.edu} \\
Observatorio Astron\'omico, Universidad Nacional de C\'ordoba,
Laprida 854, (5000) C\'ordoba, Argentina}

\date{January 10}

\pagerange{\pageref{firstpage}--\pageref{lastpage}}

\pubyear{2000}

\begin{document}

\maketitle

\label{firstpage}

\begin{abstract} We present a detailed calculation of model atmospheres for
DA white dwarfs.
Our atmosphere code solves the atmosphere structure in local thermodynamic
equilibrium with a standard partial linearization technique, which takes
into account the energy transfer by radiation and convection.
This code incorporates recent improved and extended data base of collision
induced absorption by molecular hydrogen.
We analyse the thermodynamic structure and emergent flux of atmospheres
in a range $2500 \leq T_{eff} \leq 60000$ K and $6.5\leq \log g \leq 9.0$.
Bolometric correction and colour indices are provided for a subsample of the
model grid. Comparison of the colours is made with published observational
material and results of other recent model calculations.

Motivated by the increasing interest on helium core white dwarfs, we analyse
the photometric characteristics of these stars during their cooling,
using evolutionary models recently available.
Effective temperatures, surface gravities, masses and ages have been
determined for some helium core white dwarf candidates, and their possible
binary nature is briefly discussed.

\end{abstract}


\section{Introduction}

\label{sec:intro}

White dwarf (WD) stars represent the most common end-product of the stellar
evolution. On average, WDs have a mass of $0.6 M_\odot$, a central density
of $10^6$ gr cm$^{-3}$ and a core composed of carbon and oxygen resulting
from helium burning during the preceding phases of the evolution (D'Antona
\& Mazzitelli 1990). Due to the high degree of electron degeneracy in the
interior of WDs, the outer nondegenerate layers, and specially their
atmospheres, control the loss of energy into outer space and therefore, the
cooling of these objects. From a technical view point, detailed atmosphere
models provide an adequate surface boundary condition for the computation
of evolutionary stellar models. Besides, the atmosphere models are necessary
to determine the emergent flux from the star, which is required to
interpret the luminosities and colours of observed WDs.

Current interest in WDs is motivated by two reasons. These stars are
candidates for the Galactic dark matter and can also be used to estimate the
age of the star system they belong to. The recent detection of gravitational
microlensing events in the direction of the Magellanic Glouds (Alcock et al.
1997, Alcock et al. 1999, Renault et al. 1997) suggests that very low mass
stars, WDs among others, might constitute a significant fraction of the mass
of the Galatic halo. The luminosity function necessary to study this
interpretation requires the calculation of cooling tracks using accurate
atmosphere models.

The spectral evolution of a WD during its cooling can be used to calculate
the age of the star. The first detailed calculation of WD evolution was
performed by Iben \& Tutukov (1984), for the cooling of a carbon-oxygen WD
from the planetary nucleus stage through the stage of complete internal
crystallization. Recent works (Hansen 1999; Chabrier 1999) show that the
time scales involved, particularly in an advanced stage of the cooling
process, depend strongly on the atmosphere composition (among other factors
such as the stellar mass), which through the opacity regulates the energy
loss from the stellar interior. Thus, for instance, the strong opacity from
hydrogen molecules causes a slow cooling of the WDs with hydrogen-rich
atmosphere, resulting these objects potentially observable for times
comparable to the Hubble time (Richer et al. 2000). The technique based on
the observation of a low-luminosity cutoff has been applied to estimate the
age of the local Galatic disk (Winget et al. 1987; Wood 1992) and the age of
clusters (von Hippel, Gilmore \& Jones 1995; von Hippel \& Gilmore 2000).

The majority of WDs have pure hydrogen atmospheres due to gravitational
settling, which is efficient at removing helium and heavier elements from
atmosphere toward inner layers (Fontaine \& Michaud 1979; Muchmore 1984;
Iben \& MacDonald 1985; Althaus \& Benvenuto 2000). Convective mixing could
lead to a helium-rich atmosphere in cool WD when the mass of the superficial
hydrogen layer is small enough (Koester 1976; Vauclair \& Reisse 1977;
D'Antona \& Mazzitelli 1979). In particular, Bergeron, Ruiz \& Leggett
(1997) have reported evidence for convective mixing in cool WD atmospheres.
However, hydrogen-rich atmospheres are observationally most abundant than
those of helium or other with metals, in part due to the long cooling times.
Furthermore, these atmospheres present a structure which is insensitive to
small admixtures of helium (Hansen 1999).

In this context, recent model atmosphere calculations (Hansen 1998, 1999;
Saumon \& Jacobson 1999) suggest that WDs of very low luminosity, and
therefore very old, become bluer and not redder as they age (and their
effective temperature decreases), resulting their detection easier. This
being the case, an important number of these objects should be observed with
the new generation of telescopes.

Another important question of current interest is the evolution of WDs with
very low mass having helium cores (He WD). Such stars, due to the low
stellar mass, would be formed from some binary systems. Indeed, numerous
binary configurations of this type have been observed lately (Marsh 1995;
Moran, Marsh \& Bragaglia 1997; Edmonds et al. 1999). In particular, low
mass WDs have been detected in binary systems composed of a WD star and a
pulsar. A particularly interesting binary system with these characteristics
is the millisecond pulsar J1012+5307 and its WD companion, because it is
possible to determine the properties of the pulsar from the analysis
of the WD (Van Kerwijk, Bergeron \& Kulkarni 1996). The study of these
binaries would allow us to determine an upper limit to the neutron-star mass
(and in this way, to constrain the still uncertain properties of the equation
of state of the cool nuclear matter) and on its age (with implications for
the
evolution of its magnetic field). Obviously, to obtain accurate results from
the study of these systems it is necessary to have suitable evolution models
and, in particular, detailed model atmospheres for WDs.

The primary aim of this paper is to present results of model calculations
from an independent atmosphere code. This will be applied to determine
accurate cooling tracks of WDs, by providing appropriate outer boundary
conditions to evolution models. Our model atmospheres with pure hydrogen
compositions cover extensive ranges of effective temperature ($2500 \leq
T_{eff} \leq 60000$ K) and surface gravity ($6.5 \leq \log g \leq 9.0$).
Although several sets of models available in the literature
(cf. Bergeron, Ruiz \& Leggett 1997; Saumon \& Jacobson
1999; Hansen 1999; Chabrier 1999) have already investigated different parts
of this ($T_{eff},g)$ region, we offer here an overall analysis. We have
considered it important to extend the study to the very low gravities
(log $g=6.5$)
which are found in early stages of the WD formation, and also in the cool
stages of He WDs with very low mass, which have received little or no
attention so far. In particular, we have also computed model atmospheres
following the cooling tracks of He WDs with hydrogen envelopes and several
masses.

The results presented here are based on ideal equations of state.
The incorporation of density effects in our code using the
occupation probability formalism of Hummer \& Mihalas (1988) is in progress. 
However, nonideal effects can be neglected for hydrogen WD models with
T$_{eff} \geq 4000$ K (Bergeron, Saumon \& Wesemael 1995a). They
increase for lower T$_{eff}$ and are significant for T$_{eff} \leq 2500$
K (Saumon \& Jacobson 1999).

In Section \ref{sec:ther} we derive explicit analytical expressions for the
thermodynamic quantities required in the calculations.
The method used for the model atmosphere computations is presented in
Section \ref{sec:model}, together with the opacities and convective theory
adopted.
In Section \ref {sec:resu} we display the atmosphere structure of WDs
obtained from our code. Comparisons with observations and other recent
model results, through the analysis of colour diagrams, are also presented
there. Our conclusions are given in Section \ref{sec:concl}.

\section{ Thermodynamic quantities}

\label{sec:ther}

The computation of convective flux of energy requires some properties of the
gas such as the adiabatic gradient and the specific heat. To determine them,
we consider a mixture of H$_2$, H, H$^{+}$, e$^{-}$, and radiation occuping
a volume $V$, in thermodynamic equilibrium at temperature $T$. In the
present calculations, species with low abundances can be ignored (Saumon \&
Chabrier 1992).

Let $N$ be the number of hydrogen nuclei or protons. Then, the number
conservation and charge neutrality constrains demand that the particle
density is 
\begin{eqnarray}  \label{th1}
\frac {xN}V \;,
\end{eqnarray}
with 
\begin{eqnarray}
x=1+x_1-x_2 \;,
\end{eqnarray}
where $x_1=n_{H^+}/n$ and $x_2=n_{H_2}/n$ (with $n=N/V$) are the number
concentration of H$^{+}$ and H$_2$. We use the ideal state equation to
express the gas pressure. The total pressure, including the radiation
contribution, is then 
\begin{eqnarray}  \label{th2}
P= \frac{xNkT}V+\frac 13aT^4 \;.
\end{eqnarray}

We adopt a simple approach for the total energy in the system
\begin{eqnarray}  \label{th3}
U=\frac 32 xNkT+\left[ x_1\chi _1+\left( \frac 12-x_2\right) \chi _2\right]
N\, \cr +\ \left[ kT+kT_v\left( \frac 12+\frac 1{e^\theta -1}\right) \right]
x_2N+aT^4V \;,
\end{eqnarray}
where $\theta =T_v/T$. The first term on the right-hand side is the
translational energy, which assumes the particles to behave as classical
point particles. The second term gives the potential energy of the hydrogen
nuclei, which increases with the dissociation of H$_2$ molecules and the
ionization of H atoms according to the corresponding binding energies $\chi
_1$ and $\chi _2$. For simplicity, we ignore in this term the excited states
of these species. The zero point of the energy scale is taken to be the
ground state of H$_2$. The third term represents the rotation-vibration
energy of H$_2$, where we use a vibration temperature $T_v=6100$ K. Finally,
the last term takes into account the energy in the radiation field.

Expressions for the adiabatic gradient $\nabla _A = \left. {%
\partial \ln T / \partial \ln P}\right| _S$, and heat capacity at constant
pressure $C_P=\left. \partial U/\partial T\right| _P + %
P\,\left. \partial V/ \partial T\right| _P$,
can be obtain from differentiation of Equations (\ref{th2}) and (\ref{th3}).
We have followed a similar procedure to that of Krishna-Swamy (1961; see also
Mihalas 1965) and, therefore, we only show here the results obtained
including molecular species in the gas mixture.
The expresion for the adiabatic gradient is
\begin{eqnarray}  \label{th14}
\nabla _A=\frac{a_1}{a_2} \;,
\end{eqnarray}
with 
\begin{eqnarray}  \label{th15}
a_1=1+4\gamma -b_1\left. \frac{\partial x_1}{\partial\ln P_g} \right|
_T+b_2\left. \frac{\partial x_2}{\partial \ln P_g}\right| _T 
\end{eqnarray}
and 
\begin{eqnarray}  \label{th16}
a_2=\frac 52+16\gamma +\left[ 1+\frac{\theta ^2e^{-\theta }} {(1-e^{-\theta
})^2}\right] \frac{x_2}{x}+b_1\left. \frac{\partial x_1} {\partial \ln T}
\right| _{P_g} \cr -\ b_2\left. \frac{\partial x_2}{\partial \ln T}\right|
_{P_g} \;,
\end{eqnarray}
where 
\begin{eqnarray}  \label{th8}
b_1=\left( \frac 52+\frac{\chi _1}{kT}+4\gamma \right) \frac 1{x} \;,
\end{eqnarray}
\begin{eqnarray}  \label{th9}
b_2=\left\{ \frac 52+\frac{\chi _2}{kT}+4\gamma -\left[ 1+\theta \left(
\frac 12+\frac{e^{-\theta }}{1-e^{-\theta }}\right) \right] \right\} \frac
1{x} \;.
\end{eqnarray}
For the heat capacity we obtain
\begin{eqnarray}
\frac{C_P}{kN}=\left[ \frac 32+12\gamma +\left( 1+4\gamma \right) ^2\right] x%
\cr +\left[ 1+\frac{\theta ^2e^{-\theta }}{(1-e^{-\theta })^2}\right]
x_2+\left( \frac 32+\frac{\chi _1}{kT}\right) \left. \frac{\partial x_1}{%
\partial \ln T}\right| _{P}\cr -\ \left\{ \frac 32+\frac{\chi _2}{kT}%
-\left[ 1+\theta \left( \frac 12+\frac{e^{-\theta }}{1-e^{-\theta }}\right)
\right] \right\} \left. \frac{\partial x_2}{\partial \ln T}\right| _{P}\,.
\label{hc7}
\end{eqnarray}
The partial derivatives of the concentrations $x_1$ and $x_2$, have been
calculated considering that, except in the regime of pressure dissociation,
H$^{+}$ and H$_2$ do not have simultaneously high abundances. Then, using
Saha equations for the first and third reactions in Table 1, result
\begin{eqnarray}  \label{th12a}
\left. \frac{\partial x_1}{\partial \ln P_g}\right| _T=-\frac{x_1}2\left(
1-x_1^{\;\;2}\right) \;,
\end{eqnarray}
\begin{eqnarray}  \label{th12b}
\left. \frac{\partial x_1}{\partial \ln T}\right| _{P_g}=\frac{x_1}2\left(
1-x_1^{\;\;2}\right) \left( \frac 52+\frac{\chi _1}{kT}\right) \;,
\end{eqnarray}
\begin{eqnarray}  \label{th14b}
\left. \frac{\partial x_1}{\partial \ln T}\right| _{P}=\frac{x_1}2\left(
1-x_1^{\;\;2}\right) \left( \frac 52+\frac{\chi _1}{kT}+4\gamma \right) \;,
\end{eqnarray}
\begin{eqnarray}  \label{th13a}
\left. \frac{\partial x_2}{\partial \ln P_g}\right| _T=\left( \frac
1{x_2}+\frac 4{1-2x_2}-\frac 1{1-x_2}\right) ^{-1}\;,
\end{eqnarray}
\begin{eqnarray}  \label{th13b}
\left. \frac{\partial x_2}{\partial \ln T}\right| _{P_g} =-\left( \frac 52+%
\frac{\chi _2}{kT}\right) \left. \frac{\partial x_2}{\partial \ln P_g}%
\right| _T\;,
\end{eqnarray}
\begin{eqnarray}  \label{th15b}
\left. \frac{\partial x_2}{\partial \ln T}\right| _{P} =-\left( \frac 52+%
\frac{\chi _2}{kT}+4\gamma\right) \left. \frac{\partial x_2}{\partial%
\ln P_g}\right| _T\;,
\end{eqnarray}
where the dependence of the partition functions with the temperature and
pressure has been neglected.

Since the adiabatic temperature gradient forms the basis of the
Schwarzschild criterion for convective instability, this gradient plays a
central role in the computation of atmosphere models. Therefore, we check
our expression (\ref{th14}-\ref{th16}) by comparing with results from the
equation of state (EOS) for hydrogen developed by Saumon \& Chabrier
(summarized in Saumon, Chabrier \& Van Horn 1995, SCVH). This EOS,
considered the best available at present for hydrogen, is based on the
free-energy minimization technique (for details, see Graboske, Harwood \&
Rogers 1969; Fontaine, Graboske \& Van Horn 1977; Hummer \& Mihalas 1988)
and includes a sophisticated treatment of nonideal effects.

Comparisons between values for the adiabatic gradient obtained from SCVH and
our EOS are shown in Fig. 1. The curves correspond to models of WD
atmosphere with $T_{eff}=2500$, 4000, 6000 and 10000 K, and log $g=8$. Both
present and SCVH calculations agree quite well except for high pressures of
the model $T_{eff}= 2500$ K, where non-ideal effects can be present.
For the hottest two models, discrepancies at low pressure occur beyond the
regime where SCVH EOS can be applied ($4< \log P $(dyn cm$^{-2}$)$< 19$).
In particular, the upper layers of the $T_{eff}=6000$ K atmosphere are
dominated by atomic hydrogen (see Section \ref{sec:stru}), and a tendency to
a value 0.4 for the temperature gradient (corresponding to an ideal
monoatomic gas) is reproduced by our calculations.
A partial molecular formation (up to $n_{H_2}/n=0.12$) takes place as one
proceeds inward in this atmosphere, yielding a dramatic drop in the value
of $\nabla_A$.

Other familiar features can be identified in this figure. For instance,
$T_{eff}=2500$ K model represents an atmosphere formed mainly by H$_2$. At
low temperatures present in its superficial layers, the H$_2$ vibrational
modes remain frozen and then, the adiabatic gradient takes a value close to 
$2/7=0.286$, i.e. that due to a fluid of rigid rotating molecules with two
degrees of rotational freedom and three degrees of translational freedom.
The adiabatic gradient decreases toward deep layers where the molecular gas
is partially dissociated.

The low values reached by $\nabla_A$ in the $T_{eff}=4000$ K model are due to
a marked sensitivity of the molecular abundance to changes in temperature
and pressure. While the behaviour of $\nabla_A$ in the hottest model
($T_{eff}=10000$ K) is produced by different ionization degree at
different depth.

In conclusion, the relative simplicity of our EOS allows us to obtain the
analytic expressions for the heat capacity and the adiabatic gradient
necessary to compute atmosphere models. This has the advantage of avoiding
potential errors due to numerical inaccuracies that can be present in a
sophisticated EOS, which requires interpolation from tables and numerical
evaluation of derivatives.

\section{ Model atmosphere}

\label{sec:model}

Cooling of WDs is a very slow process so that the atmosphere structure
responds to quasi-static boundary conditions. Furthermore, the region where
the emergent radiation is formed has a thickness much less than the stellar
radius. These properties allow us to choose a static model atmosphere with
the approximation of plane-parallel layers and constant surface gravity
throughout the atmosphere.

On the other hand, because the gas densities are very high in the layers
where the emergent radiation is formed, local thermodynamic equilibrium
(LTE) is a good approximation for the matter state and we adopt that.

We take the atmophere gas to be composed of H, H$_2$, e$^{-}$, H$%
^{-}$, H$^{+}$, H$_2^{+}$ and H$_3^{+}$. The main constituents are H$_2$,
H, H$^{+}$ and e$^{-}$, depending on the physical conditions. The other
species are usually present as traces, but they can play an important role
in the absorption and emission of radiation. In fact, the H$^{-}$ ion can
be a dominant source of opacity in cool atmospheres, whereas H$_2^{+}$ and
H$_3^{+}$ are important because their presence affects the concentration
of H$^{-}$ and e$^{-}$.

\begin{table}
\caption{ Reactions considered at the chemical equilibrium.}
\label{tbl-1}
\begin{tabular}{lr}
Reaction & ionization/dissociation energy (eV) \\ \hline
H $\rightarrow$ H$^+$ + e$^-$ & 13.598 \\ 
H$^-$ $\rightarrow$ H + e$^-$ &  0.755 \\ 
H$_2$ $\rightarrow$ 2H        & 4.478 \\ 
H$_2^+$ $\rightarrow$ H + H$^+$ & 2.651 \\ 
H$_3^+$ $\rightarrow$ H$_2$ + H$^+$ & 4.354 \\ \hline
\end{tabular}
\end{table}

To determine the equilibrium concentrations we consider the reactions given
in Table 1. The values of energy are those compiled by Lenzuni \&
Saumon (1992). We use the partition functions calculated by Irwin (1981) for 
H and H$_2$, by Sauval \& Tatum (1984) for H$_2^+$, and Neale
\& Tennyson (1995) for H$_3^+$.
We have checked our chemical equilibrium against results from an independent
code (Saumon 1999; based on an equation of ideal state, except in a sharp
cuttoff in the H partition function depending on the distance to the nearest
neighbor). Though some differences (at most 12\%) appear between both
calculations (mainly due to different ways of computing the H$_2$ partition
function), the agreement is satisfactory in general.

\subsection{ Opacities}

\label{sec:opa}

We have consider the most important opacity sources in a hydrogen atmosphere.
Bound-free and free-free absorption coeficients of H have been taken from
Mihalas (1978), with Gaunt factors from Menzel \& Pekeris (1935).
Isotropic Thomson scattering is assumed for the opacity of free electrons.
We adopted absorption coeficients of the negative hydrogen ion according
to John (1988) for bound-free transitions, and analytic fits by Gray
(1992) based on Bell \& Berrington (1987) results for free-free transitions.

Collisions between H$_2$ molecules yield collision-induced absorption (CIA),
consisting of rototranslational and rotovibrational bands with
peaks located in far-red and infrared spectral regions. We use H$_2$-H$_2$
CIA cross sections calculated by Borysow and coworkers, including the most
recent results (Borysow, Jorgensen \& Zheng 1997).
For increasingly large densities, absorption from collisions involving more
than two H$_2$ molecules becomes important. Three-body collisions are no
longer negligible for densities $\rho >0.02$ g cm$^{-3}$ and they dominate
the opacity above $\rho >0.2$ g cm$^{-3}$ (Lenzuni \& Saumon 1992). Ignoring
terms of higher order, the total frequency-integrated CIA coefficient can be
expressed as (Lenzuni, Chernoff \& Salpeter 1991) 
\begin{eqnarray}  \label{opa1}
k=k_2\rho ^2+k_3\rho ^3 \;,
\end{eqnarray}
where $k_2$ and $k_3$ account for absorption by binary and ternary
collisions. We assume the ratio $k_3/k_2=0.05/\rho $ meassured experimentally
by Hare \& Welsh (1958).

Finally, cross section of Rayleigh scattering from molecular hydrogen
is taken from Dalgarno \& Willians (1966). We also consider other opacity
sources such as Rayleigh scattering from H, bound-free and free-free
absorption of H$_2^{+}$, and free-free opacity of H$_2^{-}$.
For all of them, we use coefficient expressions from Kurucz (1970).

\subsection{Convective Transport}

\label{sec:conve}

Energy transport by convection is treated within the mixing-length theory
(e.g., Cox \& Giuli 1968; Mihalas 1978) in
most model atmosphere calculations of WDs. Although this phenomenological
formalism involves free parameters, Bergeron, Wesemael \& Fontaine (1992)
have found that model results with $T_{eff}$ below $\approx$ 8000 K are
insensitive to the assumed parameterization. In addition, they found that the
predicted emergent fluxes show dependence on the efficiency of convection in
the range $T_{eff}$ $\approx$ 8000-15000 K. However, it is difficult to
decide which is the most appropriate parameterization for WDs. Preliminary
analyses of the effects of convective efficiency on atmospheric parameters
of ZZ Ceti \footnote{%
WDs of DA class (i.e. with Balmer lines and no He I or metals present, and
therefore, having H-rich atmospheres) showing non-radial pulsations.} and
other lukewarm DA stars, suggest that ML1 and ML3 parameterizations
(following the nomenclature of Fontaine, Villeneuve \& Wilson 1981) could be
inadequate (Wesemael et al. 1991, Bergeron et al. 1992). In this context,
the ML2 version of the mixing-length theory has been considered a viable
alternative. The results presented here use this version of convection.

\subsection{ Numerical Method}

\label{sec:numet}

The model atmosphere may be calculated using an iteration procedure over
linearized structure equations in terms of pertubations of the temperature
and density (or pressure). However, we adopt a linearization method for the
temperature alone (Gustafsson 1971), the density and pressure structure
being obtained from a given temperature distribution by integration of
the hydrostatic equilibrium equation.
This procedure yieds similar results to a complete linearization method, but
has a much shorter computing time (Saumon et al. 1994; SBLHB).
Additional economy of execution time is obtained from formulating the
radiative transfer in terms of moment equations, with closure using variable
Eddington factors (Auer \& Mihalas 1970).

To calculate a model atmosphere, the equations of radiative transfer and
constant flux condition are linearized as functions of the temperature.
The equations are discretized and organized following the Rybicki scheme
(for details see Gustafsson 1971 and Gustafsson \& Nissen 1972).
The calculation of a model starts with a given temperature distribution from
which the hydrostatic equilibrium equation is integrated, and the opacities
and thermodynamic quantities derived. Then, transfer and energy constrain
equations are solved and we obtain a temperature correction.
The whole process is iterated to convergence. The Eddington factors are
evaluated in each step from a Feautrier solution of the radiative transfer,
using known values of the source function.

\subsection{ Model convergence properties}

\label{sec:prop}

We remark on some points about the performance of the calculations.
Although the correction procedure adopted here has
proved effective and stable, certainly it is not as stable as when the
energy transfer is purely radiative. Numerical difficulties can appear to
solve the coupled equations of transfer and conservation energy when both
convective and radiative modes are present (Gustafsson 1971; Grenfell 1974;
Koester 1980; Bergeron et al. 1991; SBLHB). These difficulties are not
exclusive to partial or complete linearization methods, since they have also
been reported by Mihalas (1965) using a modification of the
temperature-correction method proposed by Avrett and Krook (1963).

We found that numerical instability can occur mainly in atmospheres of WDs
with low T$_{eff}$ and high gravity. In these models, the convection becomes
extremely efficient and, therefore, the temperature gradient is nearly
adiabatic ($\nabla-\nabla_A < 10^{-3}$ in convective deep layers). Then, the
temperature correction oscillates yielding a large convective flux in one
iteration, and zero convection in that following for the same layer,
according to the Schwarzschild instability criterion. Among the different
ways that have been proposed to stabilize the iterations (see above
references), we found convinient a similar method to the one suggested by
Koester (1980). This consists of setting $F_c=F- F_{rad}$ instead of $F_c=0$
when $\nabla < \nabla _A$ for a layer which would be convective in the
converged model.

On the other hand, a good initial temperature distribution decreases the
number of iterations necessary and increases the range of convergence of
the model in the ($T_{eff}$, log $g$) plane. For purely radiative models,
one can start with an approximate solution of the gray atmosphere problem,
e.g. the Eddington distribution (Mihalas 1978).
However, when the convection is present, this class of distributions usually
leads to an enormous convective flux which may have destabilizing
consequences on the temperature correction method. A more suitable starting
model is obtained when in the convective zone where $F_c > F$,
the temperature gradient is changed and
forced to yield $F_c = F$ (Nordlund 1974). This procedure yields a suitable
initial temperature stratification in convective models, from which to apply
the rigorous temperature correction scheme.

Finally, as in the report of SBLHB, we found that a discontinuity
develops in the temperature distribution of converged models with T$_{eff}
\approx $ 4400-5500 K at $\log g\ga7.5$, extending to lower effective
temperatures when the gravity decrease. Such as discussed further by
Bergeron et al. (1995a), this effect is due to competition between opacities
of H$_2$ molecule and hydrogen negative ion. This convergence problem can be
solved using an appropriate form of the radiative equilibrium equation
(SBLHB), but unfortunately needs a large number of iterations.

\section{Results and discussion}

\label{sec:resu}

We have calculated a grid of models covering an extensive range of effective
temperatures $2500 \leq T_{eff} \leq 60000$ K and surface gravities $6 \leq
\log g \leq 9$. In these computations we used 76 depth points to solve the
hydrostatic equilibrium and energy tranfer equations in the range $-6 \leq
\log \tau \leq 1.5$. The continuum radiation field was computed at 87 to 107
frequency points depending on the effective temperature of the model. The
points were also chosen taking into account the location of photometric
bandpasses used in the colour computations (see Section \ref{sec:colour}).

The temperature-correction procedure was repeated until the temperature
changes were everywhere $\Delta T/T < 0.1 \%$, for most models. In hot
models, $T_{eff} \ga 14000$ K, which are radiative or weakly convective,
less than 10 iterations were sufficient to reach the convergence, yielding
deviations from flux constancy less than 0.1 per cent everywhere. For
lukewarm models, $9000 \la T_{eff} \la  14000$ K, convergence of $\Delta F/F
< 1 \%$ is obtained with about 10 iterations. As the convection becomes
more efficient, $T_{eff} \la  9000$ K, the number of iterations increases to
about 40. In the range of temperatures where the onset of the H$_2$
formation takes place, we need to arise the iteration number to
approximately 150, due to the change of the boundary conditions in energy
balance equations (see SBLHB).

\subsection{ Atmospheric structure}

\label{sec:stru}

Strongly convective flux transport develops in the inner layers of cool and
lukewarm WD atmospheres. This occurs preferentially in zones where hydrogen
is partially ionized or dissociated, which yields a significant drop in the
value of the adiabatic temperature gradient favouring the convective
instability. Fig. 2 shows the fraction of flux carried by convection in
atmospheres with surface gravities log $g=6.5$ and 8.0. It is also shown for
log $g=6.5$ models the regions where H$_2$, H and H$^+$ in turn dominate the
gas. Convection is absent from the atmosphere in the hotter early stages of
a WD. According to our calculations, a superficial convection zone appears
as the star cools below $T_{eff}\approx12000$ K for WDs with low surface
gravities (log $g=6.5$) and below $T_{eff}\approx16000$ K for more massive
stars (log $g=8$). The convective zone extends over the region where part of
the emergent continuum radiation is formed ($0.1\la\tau\la1$).

For cooler models, approximately $T_{eff} <10000$ K for log $g=6.5$ and $%
T_{eff} <11000$ K for log $g=8.0$, the efficiency of convection increases
and a considerable fraction of the flux (more than 90 per cent) is carried
by convection in the innermost layers ($\tau \ga 10$). Radiation always
dominates in the optically thin layers $\tau\la 1$. However, thanks to high
densities and molecule formation, a little convective flux persists up to
very transparent layers, $\tau\approx 10^{-3}$, in atmospheres around $%
T_{eff}\approx 4000$ K. The computed models predict a reduction of
convection for very cool atmospheres $T_{eff} \approx 2500$ K, but this
result may be affected by non-ideal contributions to the EOS. Slight
pressure dissociation of H$_2$ can be expected at large depths in these
models, affecting the profile of the convective flux as consequence of
changes in the adiabatic gradient (Saumon \& Jacobson 1999).

For log $g=8.0$ models, the convection structure  displayed in Fig. 2 
is very similar to that showed by Bergeron et al. (1991, Fig. 2) in
the common region ($T_{eff}\ge 5000$ K). As a difference, our results show
as the molecular formation (not included in that work) strongly affects this
structure below $T_{eff}=6000$ K, increasing the convection toward
superficial layers.

Results about the run of several physical variables are shown in Figs 3-5
for selected models with different effective temperatures at log $g=6.5$ and
8.0.
For the common models at log $g=8$, a good agreement is obtained from our
$T(P)$ profiles compared to those by Bergeron, Wesemael \& Fontaine (1995a,
Fig. 2, $T_{eff}=4000$ - 10000 K) and
Saumon \& Jacobson (1999, Fig. 1, $T_{eff}=2500$ and 3000 K).
The only differences are found in the transparent layers of hot models
($T_{eff}=8000$ - 10000 K), where the line blanketing (considered by
Bergeron et al.) yields there slight changes in the temperature
distribution.

Our sets of models with log $g=6.5$ and 8.0 present similar aspects except
that higher pressures and densities are found in each model atmosphere with
increasing $g$ (Figs 3-5). In terms of $T_{eff}$, it is possible to
distinguish roughly three classes of atmospheric structures.
One corresponds to hotter models, $T_{eff}\ga 12000$ K at log $g=6.5$ and $%
T_{eff}\ga 14000$ K at log $g=8.0$, which are purely radiative or with a
superficial zone weakly convective. They show rather smooth temperature and
density gradients without additional significant details. Throughout these
atmospheres, most material is ionized.

As $T_{eff}$ decreases, a new atmospheric structure takes place. The models
with $T_{eff}\approx$ 4500-12000 K at log $g=6.5$ and $T_{eff}\approx$
4500-14000 K at log $g=8.0$, cover the range from the onset of convection
with little flux transported by it to the models where the molecular
formation becomes important. In the present group of models, the
relationship between temperature and pressure exhibits (Fig. 3) a clear
transition from a very small gradient in radiative outer layers to a steep
gradient in the top of the convective zone. In contrast, the run of
density with pressure becomes relatively shallow in the transition region
between radiative and convective zones (Fig. 5). In this sense, the model
$(T_{eff},\log g)=(12000, 6.5)$ is interesting, because it shows a
slight inversion of the density in approximately $\log P=4.7$. It can be
interpreted as result of a too fast temperature rise with increasing $P$
(see Fig. 3), which is compensated by a decrease of the
density ($P \propto \rho T$).

Finally, the coolest models studied, $2500\leq T_{eff}\la 4500$ K, show some
proper features. They are characterized by molecular formation which, as
mentioned earlier, increases the convective efficiency and extends the
convective zone toward very transparent regions (Figs 2 and 4), reaching a
maximum at $T_{eff}=4000$ K. This effect has also been found in very low
mass stars with log $g=5$ (SBLHB). At such low $T_{eff}$, our models show a
uniform temperature in the upper radiative layers and a rather sudden rise
with depth in still transparent regions (Fig 4). Deep inside the convective
zone results a low temperature gradient, which is very close to
the adiabatic value.

Fig. 5 shows that very cool atmospheres extend to extremely high density and
pressure regimes (note that the top and bottom layers of all models are
located at log $\tau= -6$ and 1.5, respectively). Non-ideal effects would
actually be very small in models with relatively low gravity log $g=6.5$,
but it can become important in the deep layers (below $\rho \approx 10^{-2}$)
of very cool models with log $g=8.0$. Nevertheless, the $T(P)$ structures
of the atmosphere models are away from the region where Saumon \&
Chabrier (1992) predict the plasma phase transition of hydrogen (Fig. 3).

Compared to hot and lukewarm models analysed above (except at $T_{eff}$
around of the onset of convection), the temperature distributions of very
cool atmospheres ($T_{eff} \leq 4500$ K) are particularly sensitive to the
value of the gravity. The surface temperature decreases and the gradient in
the ($\tau,T$) plane increases as the gravity is raised (Figs 3 and 4). This
is a consequence of the CIA opacity (Section \ref{sec:opa}) which, being a
collisional process, becomes stronger in atmospheres with high gravity
(therefore more dense) increasing the cooling of upper layers.

\subsection{ Emergent flux distribution}

\label{sec:flux}

To illustrate the behaviour of the WD spectra in the ($T_{eff},g$) plane, we
display in Figs 6a-6f the emergent flux corresponding to some selected $%
T_{eff}$ values for log $g=6.5$, 8.0 and 9.0. The importance of different
opacity sources can be noted in Fig. 7, where individual contributions to
the total monochromatic opacity are shown for some relevant physical
conditions. These correspond to a Rosseland optical depth $\tau=1$ in
atmospheres with log $g=8$ and $T_{eff}=60000$, 10000, 6000 and 2500 K (from
7a to 7d, respectively). Fig. 7d shows, with heavy lines, the opacity
contributions at $\tau=0.1$ for the 2500 K model.

Emergent fluxes for the three highest values of $T_{eff}$ clearly show the
effect of the atomic hydrogen absorption, which is a dominant source at
such high temperatures (Figs 6a-c and 7a-b). Among the spectra present in
Fig. 6, only the $T_{eff}=60000$ K models show an important radiation flux
emerging from the Lyman continuum.

At $T_{eff}=15000$ K and log $g=9$, an important convective zone (with a
maximum of 40\% of the flux transported by convection) develops between
log $\tau\approx -1.5$ and 1.3. This strongly affects the Balmer continuum
decreasing Lyman and Balmer jumps, with respect to lower gravity models (Fig.
6b). This sensitivity offers the possibility of an observational test of the
convection theory even with relatively low-quality data.

Bound-free and free-free process of H$^-$ and atomic Rayleigh scattering
start to take place as the effective temperature diminishes to 10000 K. The
most prominent feature in the spectra of these atmospheres is the Balmer
jump, which markedly weakens for large gravities. We note that these changes
in the Balmer jump can also be reproduced by the presence of helium in the
atmosphere and an increasing of its abundance (Wegner \& Schulz 1981;
Bergeron et al. 1991).

A large variety of opacity sources are found in the atmospheres of lukewarm
WDs (Fig. 7c), the H$^-$ opacity being the most important.
Since H$^-$ bound-free opacity is dominant and weakly frequency dependent,
the emergent flux of $T_{eff}=6000$ K model is relatively featureless and
resembles more a blackbody distribution (Fig. 6d). Furthermore, it is
relatively insensitive to $g$.

Fig. 7d emphasizes the importance of the molecular opacity at $\tau=1$ in a
($T_{eff},\log g)=(2500,8.0)$ model. As one proceeds outward in the
atmosphere, the opacity becomes completely dominated by CIA process and
Rayleigh scattering (heavy lines in Fig. 7d, corresponding to $\tau=0.1$),
due to a marked decrease of the abundance of all species but the H$_2$
molecule.
Besides, as the temperature decreases, separate vibrational bands are clearly
identifiable in the CIA opacity. Note that the decrease of the density
(from $\tau=1$ to $\tau=0.1$) does not affect the molecular Rayleigh
scattering contribution to the total opacity per gram, because the H$_2$
remains as the most abundant component.

Synthetic spectra obtained for $T_{eff}=2500$ and 4000 K (Figs 6e-f) show a
well-known result: the effect of CIA opacity is largest for WDs of low $%
T_{eff}$ and high gravity, corresponding to a large abundance of H$_2$ and
high density, and, therefore, a large number density of H$_2$-H$_2$ pairs.
The effect of the four vibrational bands of H$_2$ (Fig. 7d) is clearly
visible in the spectrum of models with $T_{eff}=2500$ K. This strong
absorption over infrared spectral regions coincides with the peak emission
of a blackbody at the same $T_{eff}$. As a consequence, the stellar flux is
forced to emerge at shorter wavelengths, around the minimum in the opacity
caused by the H$_2$ CIA and the Rayleigh scattering.

\subsection{ Colour indices}

\label{sec:colour}

From the emergent flux provided by the atmospheric model, we can derive
magnitudes and colours. We have calculated broadband colour indices using
the Johnson-Cousins BVRI and Johnson-Glass JHK response functions of Bessell
(1990) and Bessell \& Brett (1988), respectively. The magnitude $m$
corresponding to a bandpass with tranmission function $S^m _\lambda$ was
computed from 
\begin{eqnarray}  \label{col1}
m=-2.5\log \left[ 4\pi \left( \frac RD\right) ^2\frac{\int_0^ \infty
H_\lambda S_\lambda ^md\lambda }{\int_0^\infty S_\lambda ^md \lambda }%
\right] +c_m\quad ,
\end{eqnarray}
where $H_\lambda$ is the monochromatic Eddington flux from the model
atmosphere for a WD of radius $R$ and distance $D$ from Earth. We adopted the
calibration constants $c_m$ given in Bergeron, Ruiz \& Leggett (1997, BRL),
yielding magnitudes on the Carnegie image tube system. A set of colour
indices is obtained from the difference between magnitudes, resulting
independent of the radius and distance of the star. Fig. 8 shows the
location of the filter transmitions over two illustrative emergent fluxes.
We have also calculated the bolometric correction at the visual using the
expression derived by Bergeron, Wesemael \& Beauchamp (1995b, Eq.[3]) 
\begin{eqnarray}  \label{col2}
BC=2.5\log \int_0^\infty H_\lambda S_\lambda ^Vd\lambda -10\log
T_{eff}+15.6165 \;.
\end{eqnarray}

\begin{table*}
\begin{minipage}{90mm}
\caption{ Broadband colour indices of the log $g=6.5$ and 8.0 models}
\label{tbl-2}
\begin{tabular}{ l c c c c c c c c}
$T_{eff}$& log $g$ &$B-V$ & $V-R$ & $V-K$ & $R-I$ & $J-H$ & $H-K$ & BC($V$)\\
\hline
 4000&$6.5$&$+1.070$&$+0.689$&$+2.465$&$+0.689$&$+0.226$&$+0.058$&$-0.589$\\
 4500&$6.5$&$+0.920$&$+0.594$&$+2.436$&$+0.596$&$+0.403$&$+0.128$&$-0.470$\\
 5000&$6.5$&$+0.769$&$+0.499$&$+2.063$&$+0.501$&$+0.362$&$+0.117$&$-0.297$\\
 5500&$6.5$&$+0.632$&$+0.418$&$+1.731$&$+0.422$&$+0.324$&$+0.077$&$-0.208$\\
 6000&$6.5$&$+0.515$&$+0.351$&$+1.452$&$+0.355$&$+0.287$&$+0.045$&$-0.168$\\
 6500&$6.5$&$+0.423$&$+0.298$&$+1.211$&$+0.301$&$+0.248$&$+0.021$&$-0.161$\\
 7000&$6.5$&$+0.348$&$+0.254$&$+0.997$&$+0.255$&$+0.207$&$+0.001$&$-0.170$\\
 7500&$6.5$&$+0.282$&$+0.216$&$+0.804$&$+0.215$&$+0.169$&$-0.015$&$-0.189$\\
 8000&$6.5$&$+0.221$&$+0.180$&$+0.627$&$+0.179$&$+0.133$&$-0.029$&$-0.212$\\
     &     &        &        &        &        &        &        &        \\
 4000&$8.0$&$+1.011$&$+0.653$&$+1.778$&$+0.648$&$-0.029$&$-0.076$&$-0.398$\\
 4500&$8.0$&$+0.898$&$+0.581$&$+2.160$&$+0.583$&$+0.271$&$+0.044$&$-0.403$\\
 5000&$8.0$&$+0.767$&$+0.499$&$+2.027$&$+0.500$&$+0.341$&$+0.105$&$-0.298$\\
 5500&$8.0$&$+0.644$&$+0.423$&$+1.709$&$+0.423$&$+0.300$&$+0.085$&$-0.198$\\
 6000&$8.0$&$+0.538$&$+0.360$&$+1.430$&$+0.359$&$+0.261$&$+0.056$&$-0.154$\\
 6500&$8.0$&$+0.440$&$+0.305$&$+1.201$&$+0.305$&$+0.232$&$+0.028$&$-0.152$\\
 7000&$8.0$&$+0.362$&$+0.260$&$+1.000$&$+0.259$&$+0.200$&$+0.007$&$-0.173$\\
 7500&$8.0$&$+0.298$&$+0.223$&$+0.822$&$+0.220$&$+0.169$&$-0.010$&$-0.209$\\
 8000&$8.0$&$+0.243$&$+0.192$&$+0.660$&$+0.186$&$+0.136$&$-0.024$&$-0.246$\\
\hline
\end{tabular}
\end{minipage}
\end{table*}

In view of increasing attention on very low mass WDs in the last few
years, we report the colours and bolometric corrections for a subsample of
our grid, corresponding to low gravity (log $g=6.5$) and cool
($T_{eff}=4000$-$8000$ K) models (Table 2).
We also show the results for a canonical value $\log g=8.0$.

Fig. 9 shows the ($V-R,R-I$) two-colour diagram for all models calculated.
The solid lines connect models with same gravity and log $g=6.5$ (0.5) 9.0
(from top to botton). Also displayed is a subset of 44 DA stars (filled
circles) and 45 non-DA stars (open circles) observed by BRL and having
complete BVRI-JHK photometric data
\footnote{The sample contains WDs from the catalog of McCook \& Sion (1987)
reclassified DA by BRL.}. By comparison, we include the hydrogen atmosphere
sequence at log $g=8$ of Bergeron (2000; diamonds), and that of Saumon \&
Jacobson (1999; triangles) covering $T_{eff}\leq 4000$ K. The cooling
sequence of Richer et al. (2000; dashed line) for a WD with carbon-oxygen
core, hydrogen atmosphere and mass 0.6 $M_\odot$ is also shown.

Our models predict that all stars with hydrogen atmospheres above $%
T_{eff}\approx 4500$ K, fall on a line of almost constant surface gravity
with little dispersion. The DA stars plotted in the figure remain very close
to the synthetic colour sequences over the range $3500\la T_{eff} \la 8500$
K. In contrast, the non-DA stars reveal more dispersion, although most of
them seem to follow the hydrogen sequences. Only a few non-DA stars (marked
on the figure) show colour indices strongly affected by other non-hydrogen
sources of opacity. Among them, the very low luminosity WD LHS 3250 has a DC
classification (continuous spectrum without lines deeper than 5\%) and its
atmospheric composition is not known (Harris et al. 1999). BPM 27606 is a DQ
star (ie. with carbon features) whose spectrum presents very strong C$_2$
Swan bands (BRL). LP $790-29$ is also a DQ star.

Fig. 9 shows that the theoretical sequence at log $g=8.0$ of Bergeron (2000)
is well reproduced by our models (this occurs also for other log $g$ values,
not shown here). For the coolest atmospheres ($T_{eff}<3500$ K), our
sequence at log $g=8$ becomes bluer than that of Saumon \& Jacobson (1999).
Nevertheless, the three independent models found an equivalent location for
the turnoff in $R-I$. This turnoff is a consequence of the onset of the H$_2$
formation (which starts the CIA and molecular Rayleigh opacities), and
therefore it varies with the surface gravity. The largest values of $R-I$
correspond, at instance, to $T_{eff}=3500$ K at log $g=6.5$ and $T_{eff}=4000
$ K at log $g=9.0$. In agreement with Saumon \& Jacobson, we found that
several colour indices show the presence of a turnover at low effective
temperatures, with $B-V$ being an exception (see below).

In addition, the cooling curve of Richer et al. (2000) follows the other
sequences of models and the sample observed for $T_{eff}>4000$ K, but
crosses several sequences with different $g$ for lower $T_{eff}$ where a
constant or smoothly increasing gravity should be expected as the star
cools. We also note a certain irregular behaviour where our grid predicts
that colours are insensitive to changes in the gravity ($T_{eff}>4500$ K).

The ($B-V,V-K$) two-colour diagram is displayed in Fig. 10. There is a close
agreement between Bergeron's and our models at log $g=8.0$, except for the
hot models.
A similar turnover in $V-K$ colour (reported earlier by Mould \& Liebert
1978) is predicted by both sets of models. Nevertheless, significant
departures of our sequence from that of Bergeron can be observed below $%
B-V\approx 0.3$ ($T_{eff}>7500$ K). These discrepancies presumably have their
origin in the hydrogen line opacity and the occupation probability formalism
(Hummer \& Mihalas 1988) used by his code and not included in our
calculations.

The diagram also reveals a certain difficulty of the theory in reproducing
the observed values in the $B-V$ colour at low effective temperatures.
In fact, there is an increasing difference between observed and theoretical
colour indices above $B-V\approx 0.7$. BRL have interpreted this in terms of
a pseudo continuum opacity at the Lyman edge (not taken into account in the
atmosphere codes), which affects the flux in the $B$ magnitude (for details
see their paper).

Fig. 11 illustrates that the sample of observed stars in the ($V-I,V-K$)
diagram is well reproduced by our model grid, which is also in good
agreement with the theoretical sequence of Bergeron. It is also observed
that a marked sensitivity of $V-K$ and $V-I$ colours to the effective
temperature and surface gravity in models beyond the turnover. The reason of
this is that $I$ and $K$ bandpasses lie just where the H$_2$ CIA has strong
contributions from two major vibrational transitions, while the $V$ bandpass
receives a flux excess that is forced to emerge in the optical spectral
range. One would expect that this sensitivity in the WDs with hydrogen
atmospheres could be very useful for cosmochronological purposes.

\subsection{White dwarfs with low mass}

\label{sec:evol}

Helium core white dwarf (He WD) stars are considered the final product of
the evolution of some close binary systems. The mass of these objects should
be smaller than that required for degenerate helium ignition, $\approx 0.5$ $%
M_\odot$ (Mazzitelli 1989), thus, resulting very low mass WDs. They have
been detected in large surveys (Bragaglia et al. 1990; Bergeron, Saffer \&
Liebert 1992; Bragaglia, Renzini \& Bergeron 1995; Saffer, Livio \&
Yungelson 1998), concluding that about 10\% of presently known WDs are He
WDs.

We have used our atmosphere code to study the changes in colours and
absolute visual magnitude of He WDs during the cooling. The calculations
were carried out by using the cooling sequences of Benvenuto \& Althaus
(1998) for helium-core composition and hydrogen envelopes. The chosen
evolutionary models cover stellar masses $M=0.15$, 0.30 and 0.45 $M_\odot$
with $M_H/M=4\times10^{-3}$, $2\times10^{-3}$ and $4\times10^{-4}$,
respectively, where $M_H$ is the mass of the H envelope. We wish to point
out that recent studies (Driebe et al. 1998; Sarna, Antipova \& Muslimov
1998; Sarna, Antipova \& Ergma 1999) predict the formation of He WDs with
massive hydrogen envelopes. Since the H-burning increases with $M_H$, thus
implying longer evolutionary times, the ages estimated in this paper (using
models with rather thin hydrogen envelopes) should be considered as lower
limit values.

We summarized here the most important results. Interesting observations can
be inferred from the analysis of Fig. 12, which shows evolutionary tracks
along with an observational sample of cool stars in ($J-H,V-K$) and ($J-K,R-I
$) two-colour diagrams. The cooling tracks in both diagrams show an almost
linear behaviour following the observed photometry up to a predicted
effective temperature of approximately 4500 K. Although there is an
important scatter of observed colours, it is clear that $J-H$ and $J-K$ show
a shifting to the blue in the cool extreme of the sample, for DA and non-DA
stars. This effect is very well represented by the theoretical cooling
sequences, which gets dramatically bluer in these colour indices due to the
onset of the H$_2$ collision-induced absorption for $T_{eff} \la 4500$ K.
Note how the theoretical sequences follow the red edge in $V-K$ and $R-I$ of
the observational sample, with the less massive models on the right.

Fig. 12 seems to provide observational evidence of the presence of a strong
pressure-induced opacity in most of the cool atmospheres. It also suggests
that very cool WDs identified as non-DA, could have an important abundance
of hydrogen in their atmospheres (perhaps beyond the spectroscopic limit of
detection due to the low temperatures). A good example is LHS 1126 (see Fig.
12), the peculiar emergent flux has been interpreted in terms of H$_2$-He
CIA in a mixed H/He atmosphere with $\log N$(He)$/\log N$(H)$=0.8$ (Bergeron
et al. 1994).

Another feature exhibited in this figure is the prediction of a second
turnover in the $J-H$ colour (around $V-K=0.3$) for an advanced stage of the
cooling. In this part of the ($J-H,V-K$) two-colour diagram, the tracks of
stars with different mass coincide, but the cooling times differ
significantly. For example, the lowest value of $J-H$ is reached by a 0.15 $%
M_\odot$ He WD at 7 Gyr, while a 0.45 $M_\odot$ star needs 16 Gyr more, due
to its lower radius and, therefore, to a lower rate of cooling.

In the ($V-I,M_V$) colour-absolute magnitude diagram of Fig. 13, the tracks
corresponding to masses 0.15, 0.30 and 0.45 $M_\odot$ are shown from right
to left. The diagram shows that when a 0.15 $M_\odot$ He WD is older than 5
Gyr, it gets bluer in the $V-I$ colour. There, the absolute magnitude
changes slowly and the star remains brighter than $M_V=16.5$ up to 10 Gyr.
Similar behaviour is found for more massive models, but a longer time
(around 20 Gyr at $M=0.45$ $M_\odot$) is spent by these stars before the
turnover in $V-I$ takes place, as a consequence of an abundant H$_2$
formation in their atmospheres.

\begin{table*}
\begin{minipage}{150mm}
\caption{ Fundamental parameters of low mass white dwarfs assuming a single
helium core star}
\label{tbl-3}
\begin{tabular}{ l l l c c l c c r}
$\;\;$WD & Name & $\;T_{eff}$ (K) & log $g$ & $M/M_\odot$ & $\sigma$ (mag)&
$M_{V\;cal}$ & $M_{V\;obs}$ & Age (Gyr) \\
\hline
$0135-052$ & L$870-2$   & 7200$\pm80$ & 7.28$\pm0.08$ & 0.30$\pm0.02$ &
0.03 & 12.41    & 12.40$\pm0.07$    & 1.8$\pm0.3$  \\
$0747+073$A& LHS 239    & 4210$\pm50$  & 7.71$\pm0.08$ & 0.44$\pm0.04$ &
0.05 & 15.65    & 15.65$\pm0.03$    & 11.8$\pm1.5$ \\
$1818+126$ & G $141-2$  & 6450$\pm200$ & 7.40$\pm0.11$ & 0.33$\pm0.03$ &
0.10 & 13.06    & 13.04$\pm0.50$    & 2.2$\pm0.1$  \\
$2048+263$ & G $187-8$  & 5250$\pm150$ & 7.46$\pm0.15$ & 0.34$\pm0.06$ &
0.12 & 14.14    & 14.12$\pm0.15$     & 4.0$\pm1.3$  \\
$2248+293$ & G $128-7$  & 5600$\pm80$ & 7.55$\pm0.07$ & 0.37$\pm0.03$ &
0.05 & 13.94    & 13.94$\pm0.19$     & 3.3$\pm0.2$  \\
\hline
\end{tabular}
\end{minipage}
\end{table*}

\begin{table*}
\begin{minipage}{150mm}
\caption{ Parameters derived assuming a binary with two identical degenerate
stars}
\label{tbl-4}
\begin{tabular}{ l l l c c l c r}
$\;\;$WD & core$\;$ $\;$ $\;$ $\;$ & $\;T_{eff}$ (K) & log $g$ &
$M/M_\odot$ & $\sigma$(mag) & $M_V$ (binary) & Age (Gyr) \\
\hline
$0135-052$ & He  & 7200$\pm80$ & 7.81$\pm0.05$ & 0.50$\pm0.02$ & 0.03 &
12.41 & 2.5$\pm0.2$ \\
           & CO & 7200$\pm80$  & 7.78$\pm0.04$ & 0.47$\pm0.02$ & 0.03 &
12.40 & 1.1$\pm0.1$ \\
$0747+073$A& CO & 4180$\pm70$ & 8.22$\pm0.16$ & 0.72$\pm0.10$  & 0.13 &
15.65      & 8.6$\pm1.0$ \\
$1818+126$ & CO & 6450$\pm200$ & 7.92$\pm0.10$ & 0.54$\pm0.06$ & 0.09 &
13.04      & 1.7$\pm0.4$ \\
$2048+263$ & CO & 5250$\pm170$ & 7.98$\pm0.15$ & 0.57$\pm0.09$ & 0.12 &
14.12      & 3.8$\pm1.1$ \\
$2248+293$ & CO & 5600$\pm80$  & 8.09$\pm0.06$ & 0.64$\pm0.04$ & 0.05 &
13.94      & 3.5$\pm0.2$  \\
\hline
\end{tabular}
\end{minipage}
\end{table*}

A number of observed WDs appear redward in $V-I$ of the 0.45-$M_\odot$
sequence. Since the stellar evolution theory predicts that single WDs with
very low mass cannot be formed within the lifetime of the Galaxy, those
stars in the diagram correspond to unresolved (or barely resolved) double
degenerate systems which appear overluminous, or single low mass WDs forming
a binary system with a much less luminous companion.

We have used our code to derive atmospheric parameters, effective
temperature and surface gravity, for five of these stars using photometric
data of BRL. Combining these with evolutionary results we have also derived
masses and ages for the stars. The effective temperature can be constrained
from atmosphere fits
to the observed colour indices. However, the surface gravity is poorly
determined by this procedure for hot stars ($T_{eff}\ga 4500$ K), because
the colours are rather insensitive to changes of $g$ in such models.
Nevertheless, it is possible to derive the gravity and the mass of a star
when its absolute magnitude is known. Therefore, we determine the effective
temperature, the surface gravity and the mass of each star, fitting
simultaneously its absolute visual magnitude and colours with a
least-squares method. The $B-V$ colour was not included in this procedure
because there is an opacity source affecting the $B$ filter, which is not
considered in the model (see Section \ref{sec:colour}). The mass-radius
relations employed in the fitting and the evolution ages were taken from
models kindly provided by Benvenuto \& Althaus (2000, private communication).
All interior models used here have hydrogen envelopes.

Treating the observed objects as single WDs, we obtain the results
summarized in Table \ref{tbl-3}. Successive columns give the WD identifier,
name, effective temperature, surface gravity, estimated mass, mean deviation
of the magnitude and colour fitting, calculated and observed absolute visual
magnitude, and age. The quoted errors have been determined as the changes
necessary to produce a mean incertainty of 2$\sigma$ provided by the fitting
procedure.

The effective temperatures determined here are on average 65 K hotter than
those previously estimated by BRL. These differences are not significant
against to the mean errors of both calibrations (120 K here and 110 K in the
BRL work). As discussed above, the determination of the $T_{eff}$ is mainly
provided by the broad-band fit, whereas the temperature uncertainties can
be significantly constrained by the absolute magnitude fit.

Our $\log g$ values are 0.1 larger than the results of BRL, and we also
obtain systematically larger mass for these stars (by $\approx 0.06$
$M_\odot$), in part due to the use of different mass-radius relations.
Furthermore, using the helium core models of
Benvenuto \& Althaus, we estimated ages about twice than the BRL
determinations, who used carbon core models of Wood\footnote{%
Evolutionary models of helium core WDs are available only recently.}. These
differences have an origin mainly in that the heat content stored per gram
in a helium core is larger than that of carbon (or carbon-oxygen) core for
fixed temperature. Consequently, He WDs remain bright for comparatively
longer times, while the C and C-O WDs cool more rapidly.

Among the five stars analysed here, only L$870-2$ is confirmed as binary.
This star is a double-lined spectroscopic binary composed
of a detached pair of DA WDs (Saffer, Liebert \& Olszewski 1988).
LHS 239 is a component of a resolved binary (e.g. Poveda, Herrera, Allen,
Cordero \& Lavalley 1994). Its companion is LHS 240, a WD classified DC9 in
the McCook \& Sion catalog (1987).
For LHS 239, BRL have found a discrepancy between the effective temperature
derived from its energy distribution (colour-fit) and that obtained from
spectroscopy (fit to H$\alpha$ line profile). They consider that it could be
explained if this object is itself an unresolved double system.
The remain stars in the sample (G$141-2$, G$187-8$ and G$128-7$)
are suspected double degenerate binary (BRL)
because their derived masses fall below 0.4 $M_\odot$ and the current
theories attribute to such WDs a binary origin.

Table \ref{tbl-4} shows the results of interpreting this group of stars
as unresolved binaries containing two identical WDs.
The combined mass estimated for these systems is below the Chandrasekhar
limit, although the individual stars are too massive to have helium cores,
except perhaps in the case of L$870-2$.
It is worth noting that combinations of two different atmospheres
(not analysed here) could allow mixed systems of He WD + CO WD. The
effective temperatures in Table \ref{tbl-4} are equivalent to those deduced
from a single WD fitting due to the lack of sensitivity to gravity
variations of colours, except for LHS 239.

From a detailed spectroscopic analysis Bergeron, Wesemael, Liebert \&
Fontaine (1989) have deduced that L$870-2$ is composed of a pair of DA
WDs with comparable effective temperatures ($T_1=7470$ K and $T_2=6920$ K)
and masses ($M_1=0.47$ $M_\odot$ and $M_2=0.52$ $M_\odot$). For these
determinations, they adopted the zero-temperature mass-radius relation
of Hamada-Salpeter (1961).  Assuming identical components for the system and
using both helium core and carbon-oxygen core models recently available,
we confirm that their individual masses are within the core-helium ignition
limit ($\approx0.45-0.50$ $M_\odot$, see Table \ref{tbl-4}).
Our result for the effective temperature ($T_{eff}=7200$ K) is intermediate
between those deduced by Bergeron et al..

We note that LHS 239 cannot be modelled satisfactorily by a binary with
identical components (the mean deviation of the fit, $\sigma=0.13$, is
much larger than the uncertainty of the observed magnitude,
$\sigma_{obs}=0.03$),
while a good fit is obtained considering this star as a single object
($\sigma=0.05$). These results suggest that the photometry observed for
LHS 239 corresponds to a single He-core WD. We consider that another origin,
different to unresolved binary possibility, would be need to explain the
discrepancy in the calibrations of $T_{eff}$ obtained by BRL (see above).

The photometric fits for G$141-2$ and G$187-8$ do not show any clear
preference between the single WD or unresolved binary interpretations. In
particular, we have not found a very satisfactory hydrogen atmosphere fit to
the observed colour indices for these stars, perhaps due to a significant
presence of helium or metals in their atmospheres.
For G$128-7$, a good fit is obtained for both single and double star
possibilities. Unfortunatly, for the deduced effective temperature (5600 K),
the energy distribution is gravity-insensitive and the fits do not
highlight the binary nature of this star.
Additional spectroscopic data and line calculations could improve the
energy distribution fits and the identification of the components of
these and other (suspected) unresolved binaries.

\section{Conclusions}

\label{sec:concl}

We have computed a grid of hydrogen model atmospheres for white dwarfs
covering an extensive region in the ($T_{eff},g$) plane. For this, we have
employed a partial method of linearization of equations for stellar
atmospheres in which the energy is transported by radiation and convection.
In view of numerical instabilities usually found to solve the resulting set
of equations, we use a suitable initial temperature distribution which
improves the convergence properties taking into account the low values of
temperature gradient expected in deep layers of convective models.

From a relatively simple thermodynamic model, we have derived analytical
expressions for the adiabatic temperature gradient and heat capacities of a
mixture of hydrogen species. They result appropriate and easy to apply in
the calculation of the convective flux. In particular, for the density
regimes present in most of the atmospheres studied, our expression of the
adiabatic gradient successfully reproduces the values of the ideal equation
of state developed by Saumon \& Chabrier.

We have analysed in detail the structure and emergent flux of hydrogen white
dwarf atmospheres over a wide range of effective temperatures and gravities.
This analysis complements the most recent studies, to provide an overall
picture of these atmospheres and the changes to them expected during the
cooling. In particular, we find strong modifications of the emergent flux
for $T_{eff}\la 4500$ K produced by the pressure-induced absorption of
molecular hydrogen. The results obtained from our code have been compared to
those recently published. Our calculations confirm the importance of infrared
photometry for the study of cool white dwarfs, and its use in the
determination of ages of the Galaxy and globular clusters.

Spectral evolution of helium core white dwarfs has briefly been studied
using evolutionary sequences recently available for these objects.
Colour-colour diagrams show that the theoretical cooling sequences follow
the low mass boundary at the cool extreme of an observational sample of
white dwarfs. It is very interesting to note that on a timescale less than 9
Gyr, very low mass He WDs ($0.15$ $M_\odot$) show strongly non-blackbody
colours with absolute visual magnitude that are still below 16.5, while
massive He WDs ($\approx0.45$ $M_\odot$) remains brighter than $M_V=17$
for 25 Gyr.

Combining atmosphere calculations with interior model results, we determined
the possible fundamental parameters of some low mass white dwarfs. Because,
in general, very low mass WDs are suspected to be close binaries,
the estimated parameters depend on the possible components of each system.
With the assumption that both components are identical, the analysed stars
present preferentially carbon-oxygen core compositions but with combined
masses below the Chandrasekhar limit.

We will present self-consistent cooling calculations in the near
future, by incorporating our atmosphere code as an appropriate surface
boundary condition for evolutionary models of white dwarfs.

Acknowledgments

I am deeply grateful to O. Benvenuto and L. Althaus for encouraging me to
perform this study. I thank them for many useful and stimulating
discussions, and their thorough reading of the manuscript. L. Althaus has
provided me with invaluable help for this investigation. His suggestions
and contributions resulted in substantial improvements to the atmosphere
code and the results presented here. I am also grateful to D. Saumon, who
offered significant comments and provided me with his abundance and colour
calculations for comparison. It is a pleasure to thank P. Bergeron for
kindly making available his colour calculations and A. Borysow for providing
the updated cross-sections for the collision-induced absorption by H$_2$.
Many thanks are due to N. Cepeda and M. Gomez for kindly revising the
English text. I wish to thank also the referee for his useful remarks
which improved the earlier version of this paper.


\begin{itemize}
\item  {{\bf Figure 1.-} Comparisons of values for the adiabatic temperature
gradient $\nabla _A = \left. {\partial \ln T / \partial \ln P}\right| _S$,
obtained from SCVH EOS (dotted lines) and our expression (Eqs.
(\ref{th14}-\ref{th16}), solid lines) calculated for different atmosphere
models at log $g=8$ (with effective temperatures indicated on the plot).}

\item  {{\bf Figure 2.-} The structure of the convection zones for models
with log $g=6.5$ and $8.0$. Curves show the layers where convection carries
1, 10 (20) 90 and 95 per cent (from bottom to top) of the total flux at log $%
g=6.5$ (solid lines) and 8.0 (dashed lines). The dotted curves show from
left to right 95, 50 and 5\% of H$_2$ abundance, and 5, 50 and 95\% of H$^+$
abundance for models at log $g=6.5$.}

\item  {{\bf Figure 3.-} Temperature-pressure stratifications for atmosphere
models with log $g=6.5$ (solid lines) and 8.0 (dotted lines). Models shown
have $T_{eff}= 2500$ (250) 4000, 4500, 5000 (1000) 12000, 15000 (5000) 30000
(10000) 60000 K (from bottom to top). For models at log $g=6.5$, the
triangles indicate the $\tau=2/3$ level, and the circles show the regions
where convection carries 1\% of the total flux near the transition between
radiative and convective zones. The plasma phase transition predicted by
Saumon \& Chabrier (1992) is represented by the heavy curve labeled ``PPT''.
This figure extends the results in Fig. 2 of Bergeron et al. (1995a).}

\item  {{\bf Figure 4.-} Runs of temperature with Rosseland mean optical
depth for models presented in Fig. 3. Solid circles indicate the top of the
convective zone of models at log $g=6.5$.}

\item  {{\bf Figure 5.-} Density profiles for models as those referred to in
Fig. 4. Visible extremes of each curve are located at log $\tau= -6$
(botton) and 1.5 (top).}

\item  {{\bf Figure 6.-} Emergent flux distributions for various effective
temperatures (indicated on the plot) and log $g=6.5$ (solid lines), 8.0
(dashed lines) and 9.0 (dash-dotted lines). Those are compared with a
blackbody spectrum at the same $T_{eff}$ (dotted lines).}

\item  {{\bf Figure 7.-} In an obvious notation, individual contributions to
the opacity at a depth $\tau=1$ of the models with log $g=8.0$ and $%
T_{eff}=60000$, 10000, 6000 and 2500 K, from $(a)$ to $(d)$. These
correspond to $(\log T, \log \rho)=$ ($4.76,-6.53$), ($4.03,-5.75$), ($%
3.80,-4.15$), ($3.53,-1.13$), respectively. Heavy lines in (d) indicate the
most important opacity contributions at $\tau=0.1$ of the 2500 K model,
where $(\log T, \log \rho)=(3.09,-2.46)$.}

\item  {{\bf Figure 8.-} Normalized profiles of emergent fluxes for $%
T_{eff}=2500$ (dashed line) and 10000 K (solid line) at log $g=8$, and the
blackbody profile at 2500 K (dash-dotted line). Dots indicate the filter
transmissions of $BVRI$ Johnson-Cousins and $JHK$ Johnson-Glass systems.}

\item  {{\bf Figure 9.-} } The ($V-R,R-I$) two-colour diagram shows hydrogen
model atmospheres of white dwarfs. Heavy solid lines represent our models at
log $g=6.5$ (0.5) 9.0 from top to bottom, while thin lines connect those
with the same $T_{eff}$ (labeled). It is also shown the log $g=8.0$
sequences of Saumon \& Jacobson (1999; triangles) and Bergeron (2000;
diamonds); and the cooling track for a 0.6-$M_\odot$ carbon-oxygen WD
calculated by Richer et al. (1999; dashed line). Photometric data of DA
(filled circles) and non-DA (open circles) stars are taken from BRL, except
the colour indices of LHS 3250 which are taken from Harris et al. (1999).

\item  {\bf Figure 10.-} The ($B-V,V-K$) colour-colour diagram shows the
photometric data of BRL (circles), the log $g=8.0$ sequence of Bergeron
(diamonds) and our model grid (lines). Notation is as Fig. 9.

\item  {\bf Figure 11.-} Same as Fig. 10 for the ($V-I,V-K$) diagram.

\item  {{\bf Figure 12.-} Evolutionary tracks of He WDs with masses 0.15,
0.30 and 0.45 $M_\odot$ (solid lines from right to left) displayed in ($%
J-H,V-K)$ and ($J-K,R-I$) two-colour diagrams. A subset of cool WDs observed
by BRL is represented with filled (DA) and open circles (non-DA WDs). Ages
in Gyr are labeled for 0.15 and 0.45-$M_\odot$ models. Dashed lines indicate
the cooling track of a black body (independently of its size!).}

\item  {{\bf Figure 13.-} The colour-absolute magnitude diagram shows the
cooling tracks of WDs with helium cores, hydrogen atmospheres and masses
0.15, 0.30 and 0.45 $M_\odot$ (solid lines from right to left). Ages are
indicated in units of 10$^9$ years. Dotted lines show the same cooling
tracks assuming stellar radiation as black body. The observed photometry is
taken from BRL (DA WDs: filled circles; non-DA WDs: open circles); Monet et
al. (1992; diamonds) and Harris et al. (1999; LHS 3250).}
\end{itemize}

\bsp

\label{lastpage}

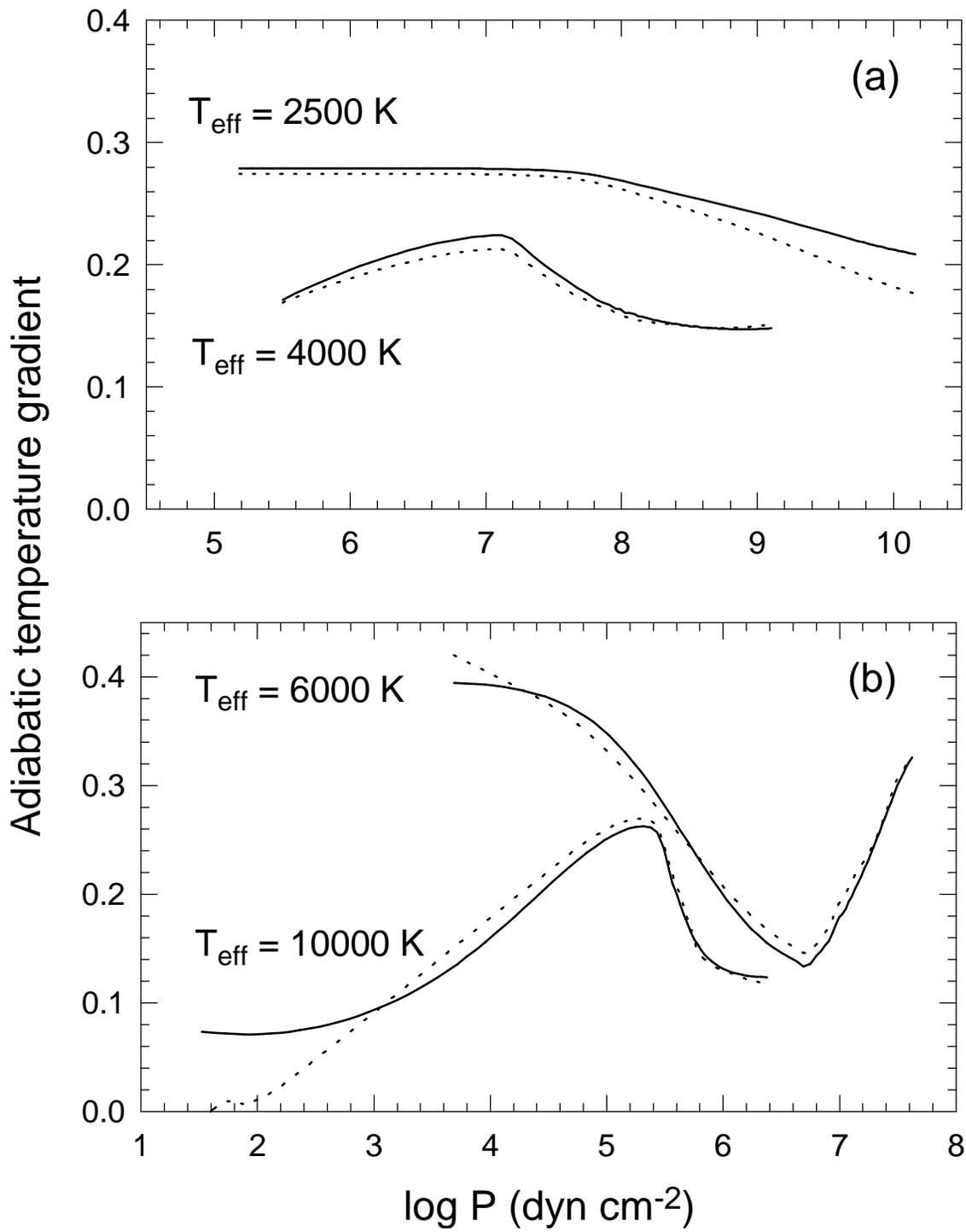
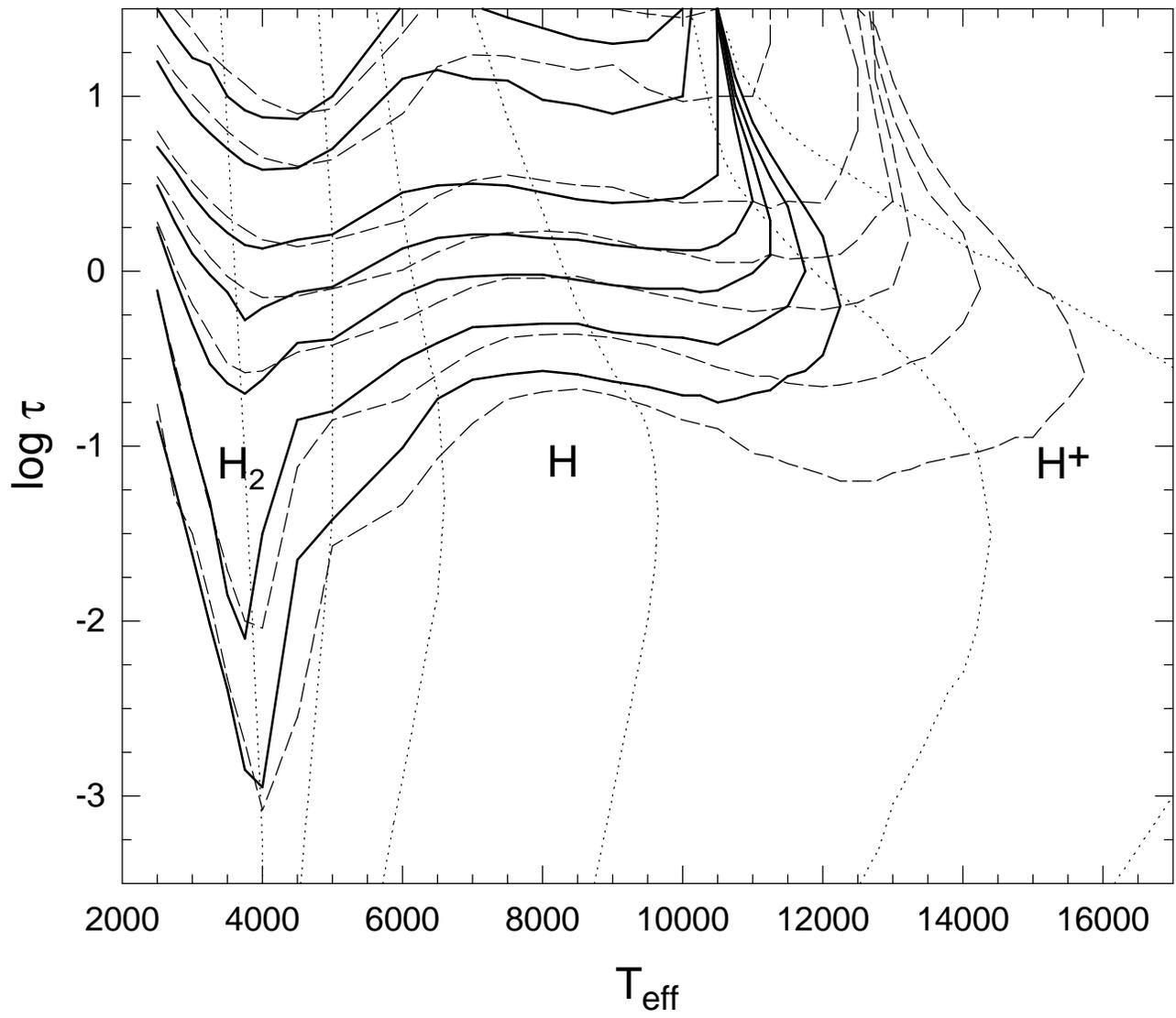
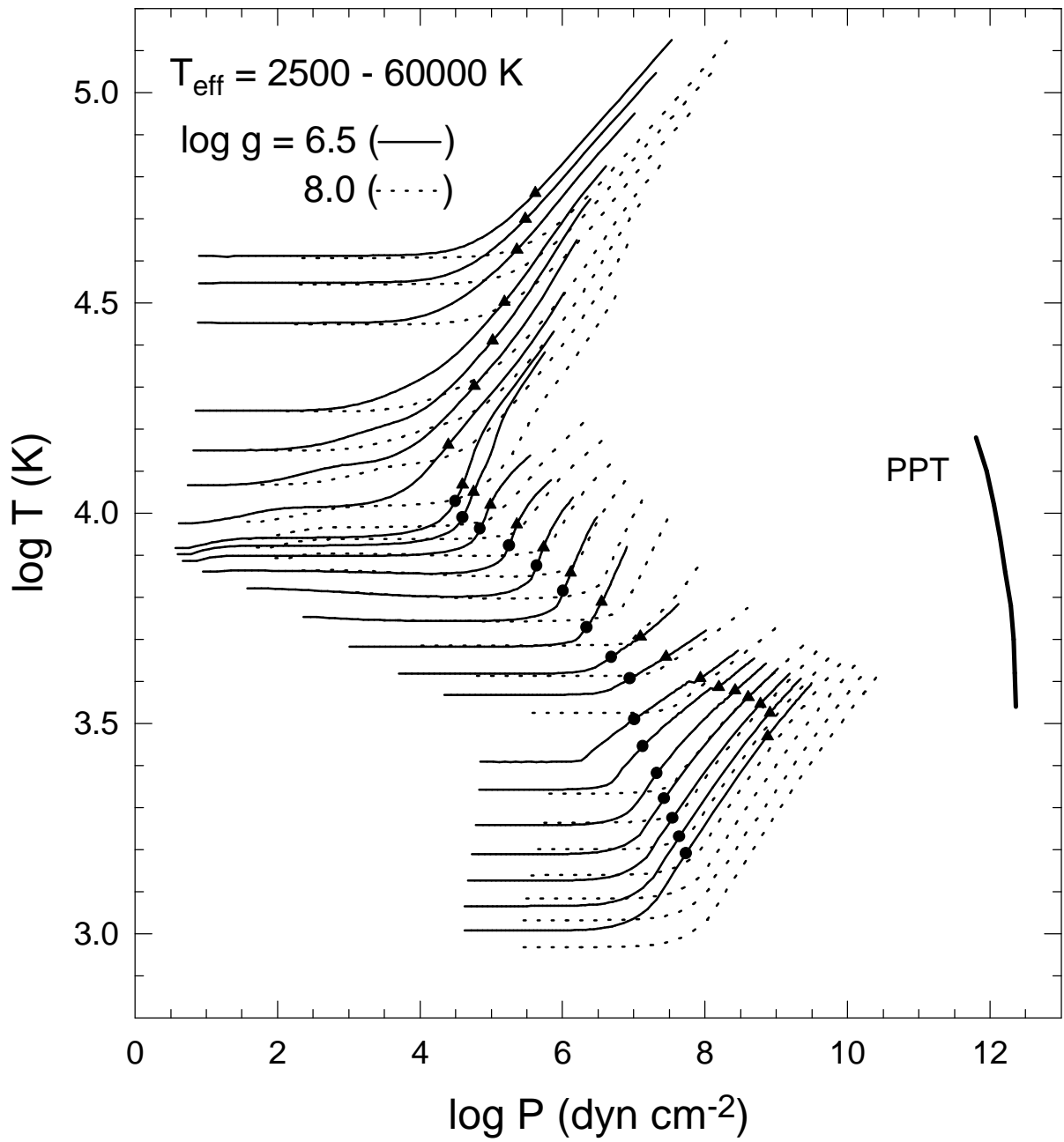
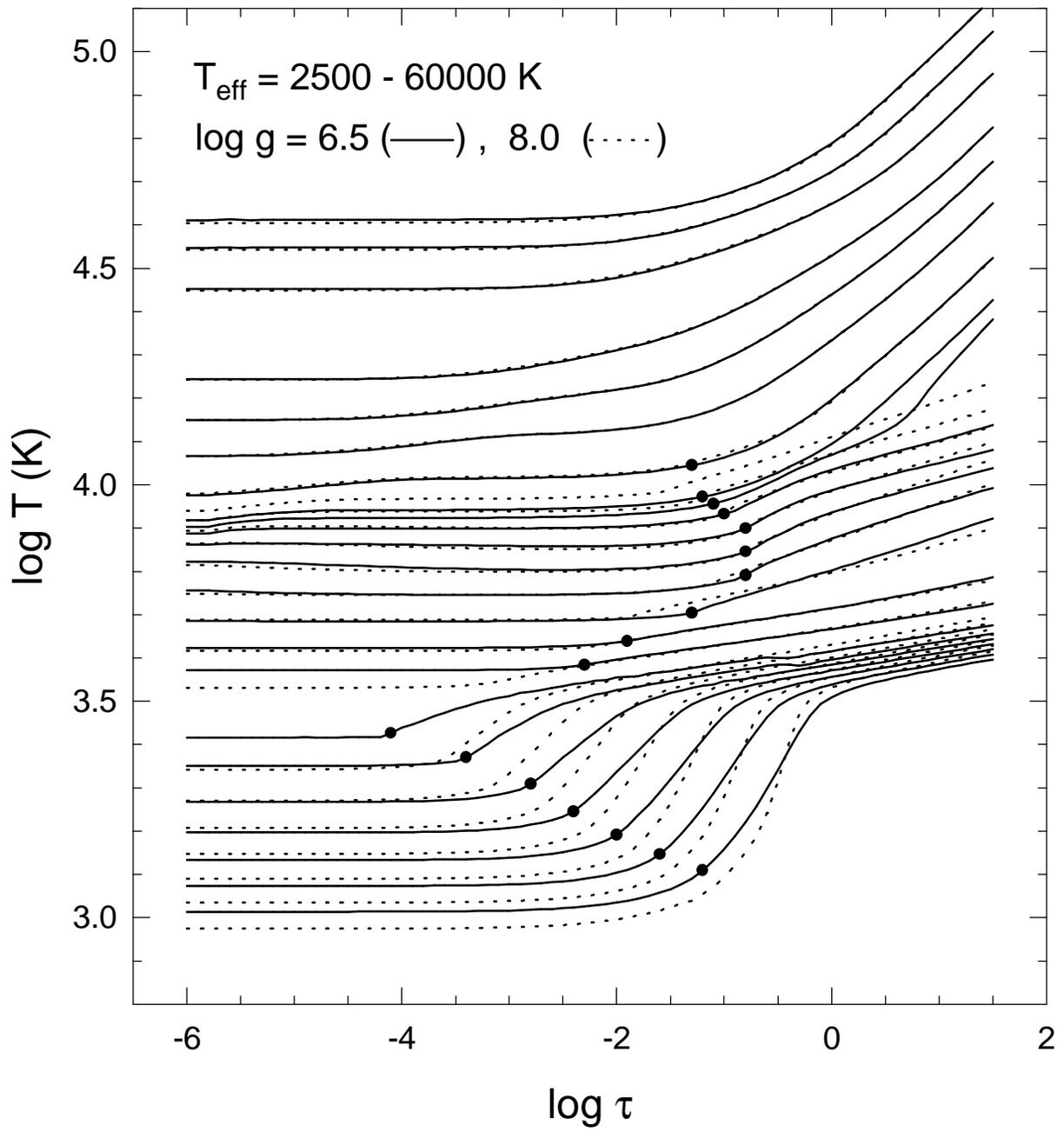
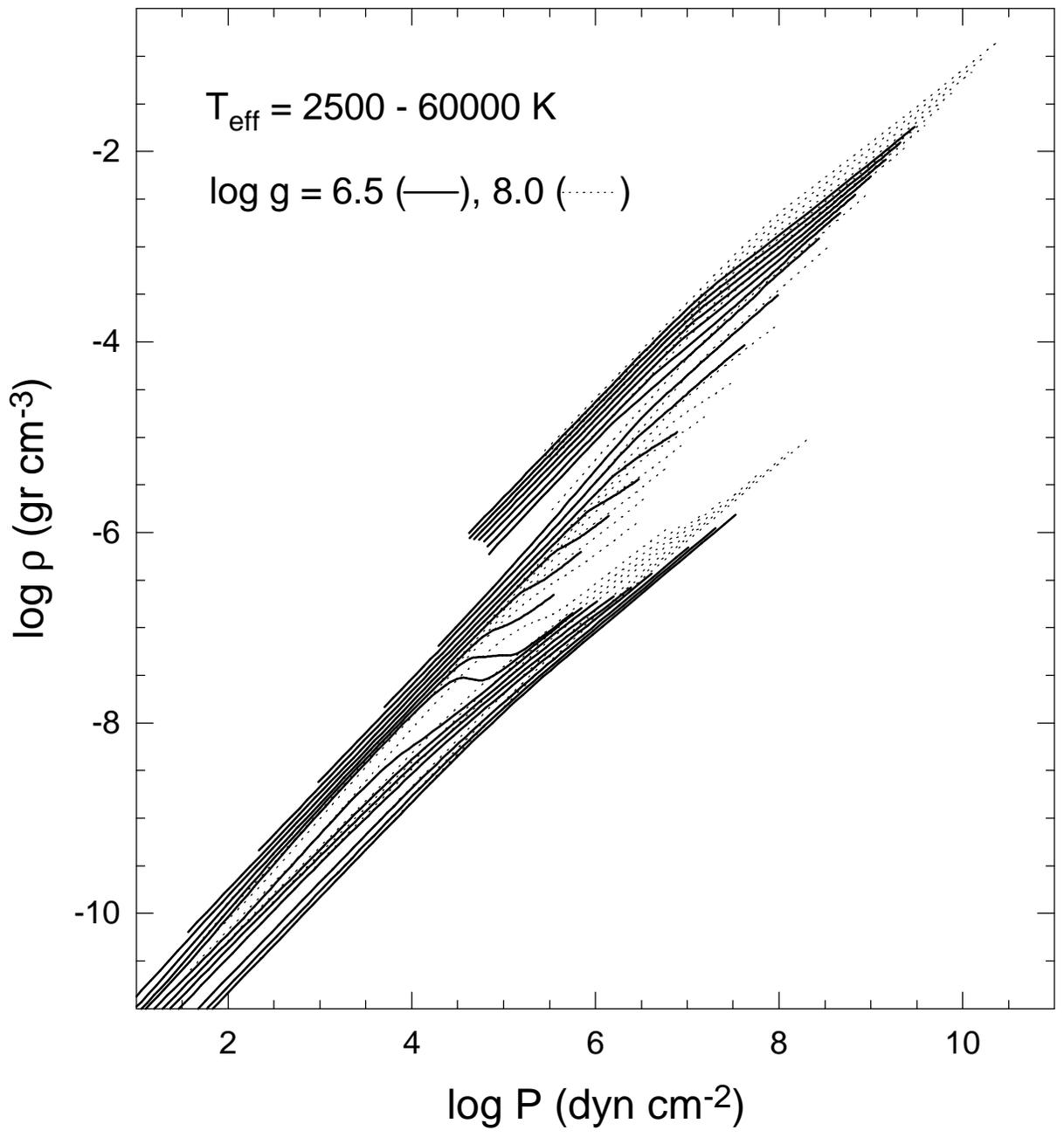
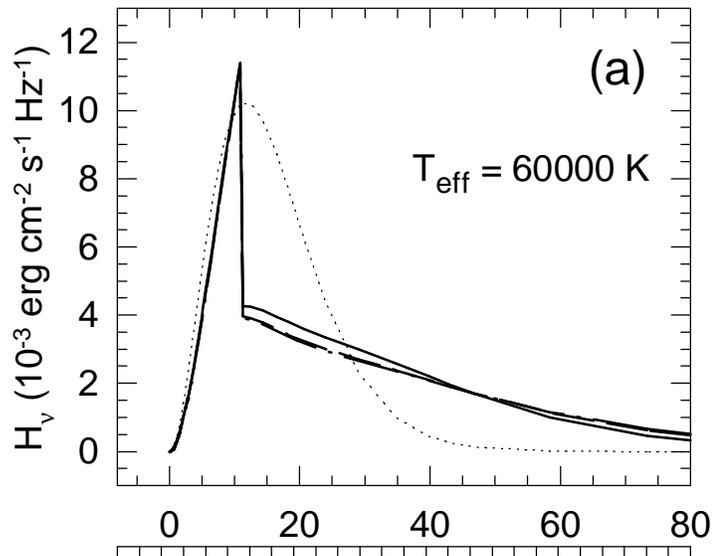
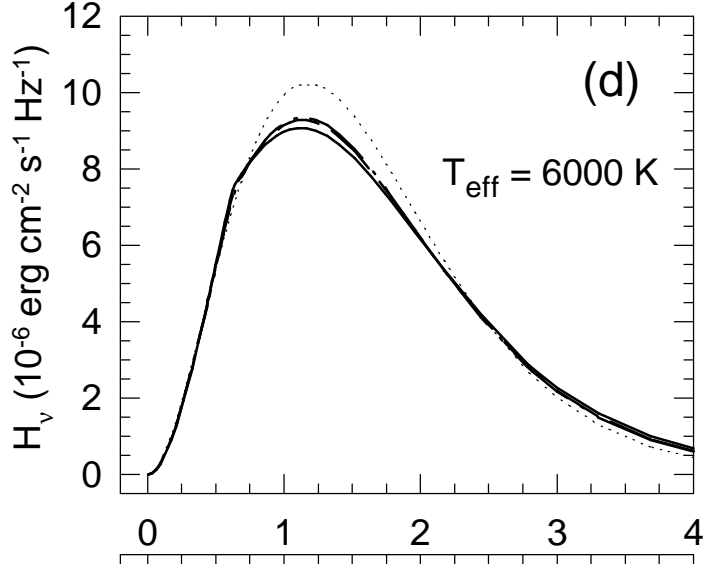
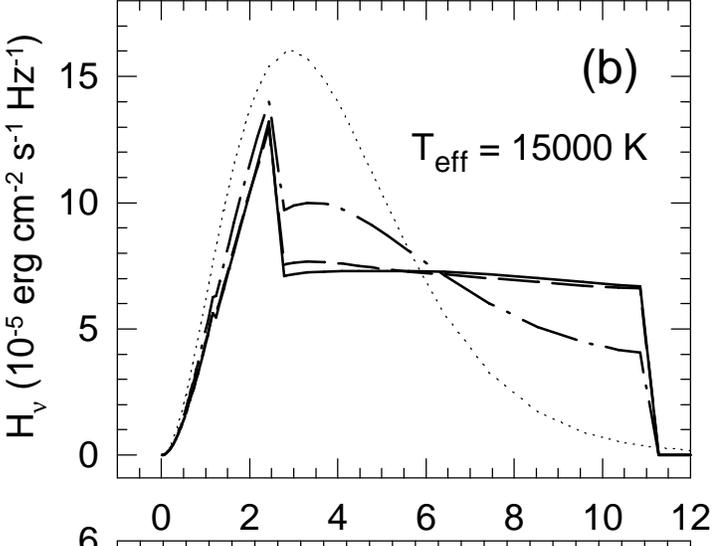
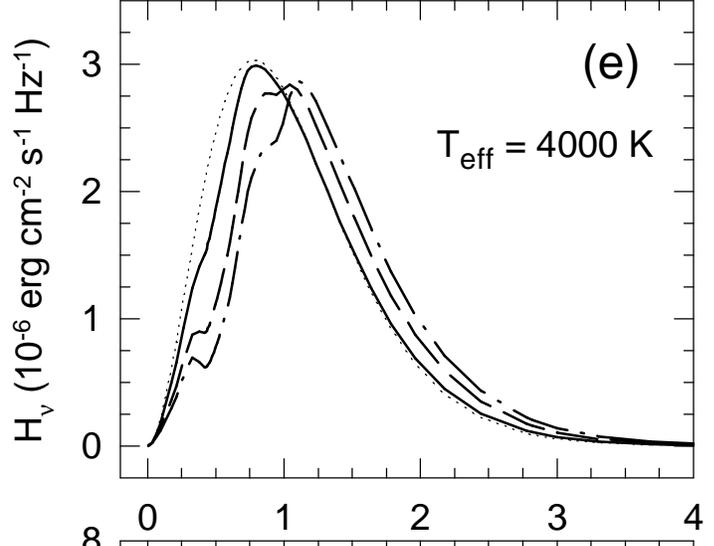
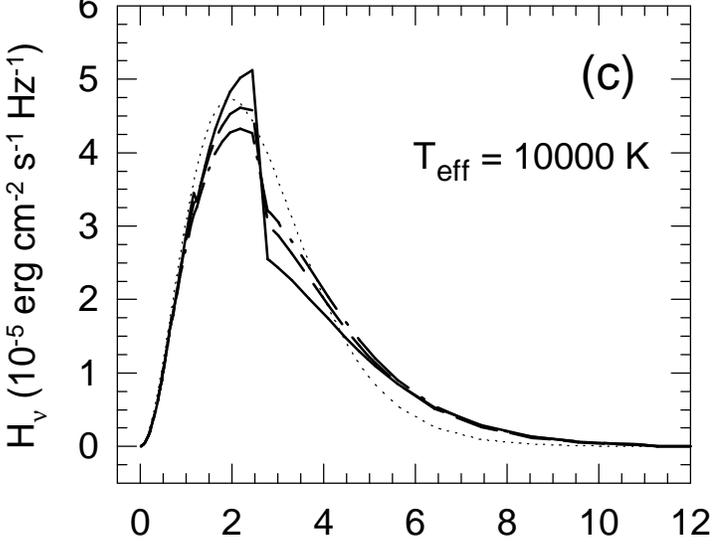
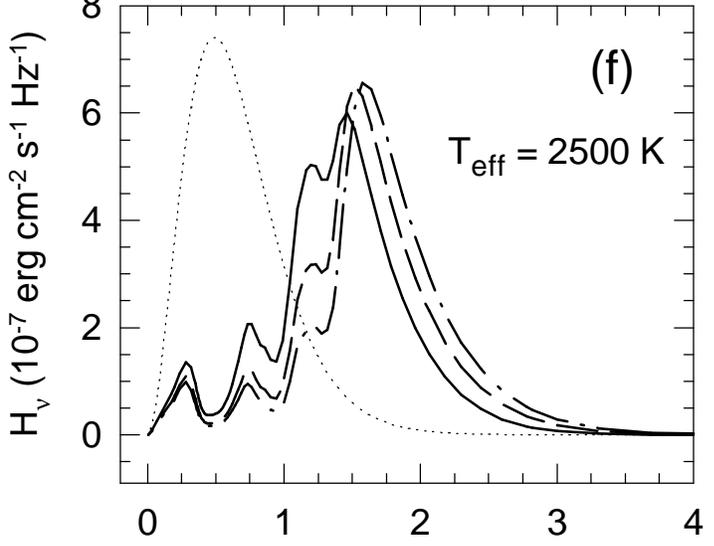
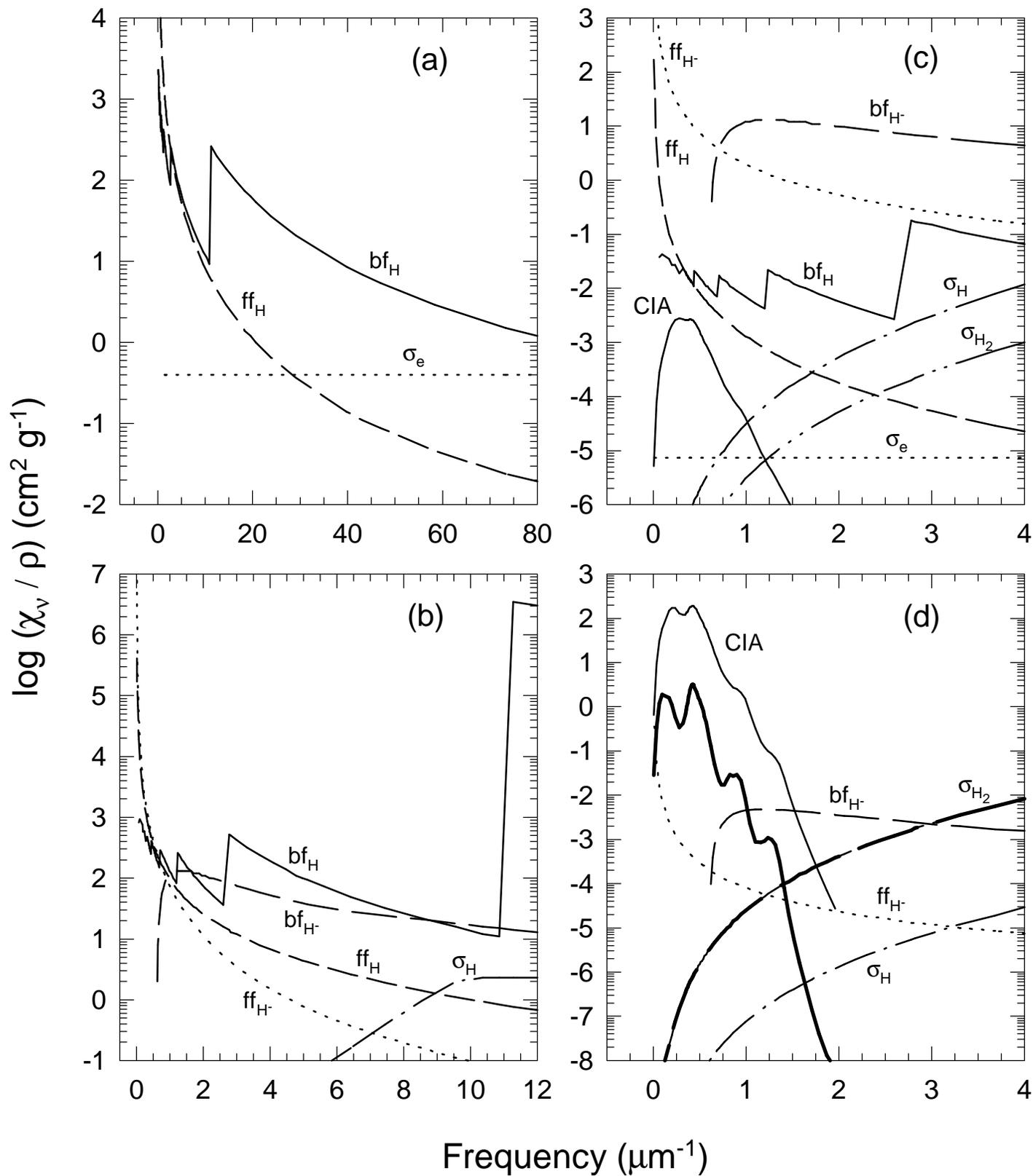
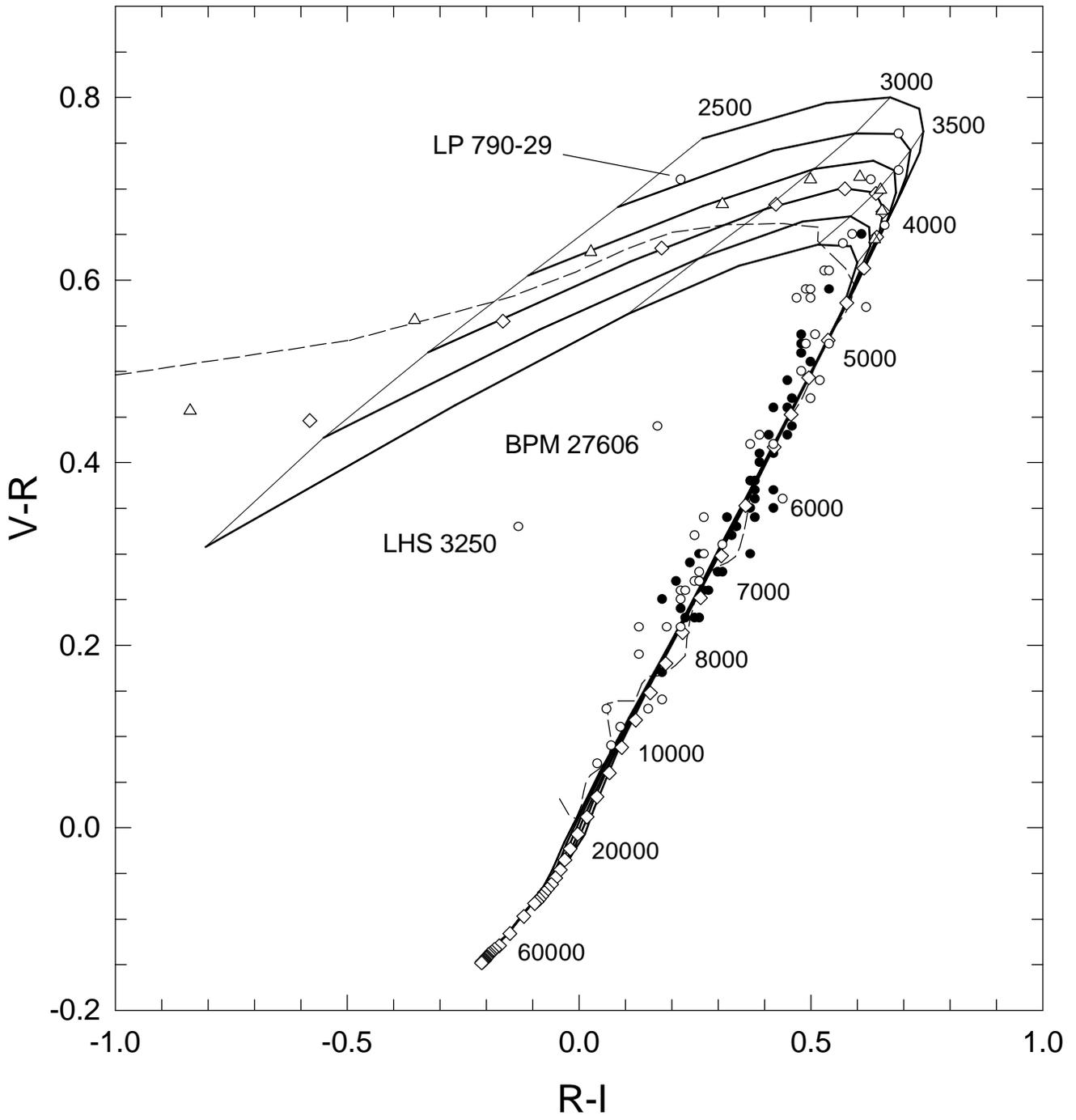
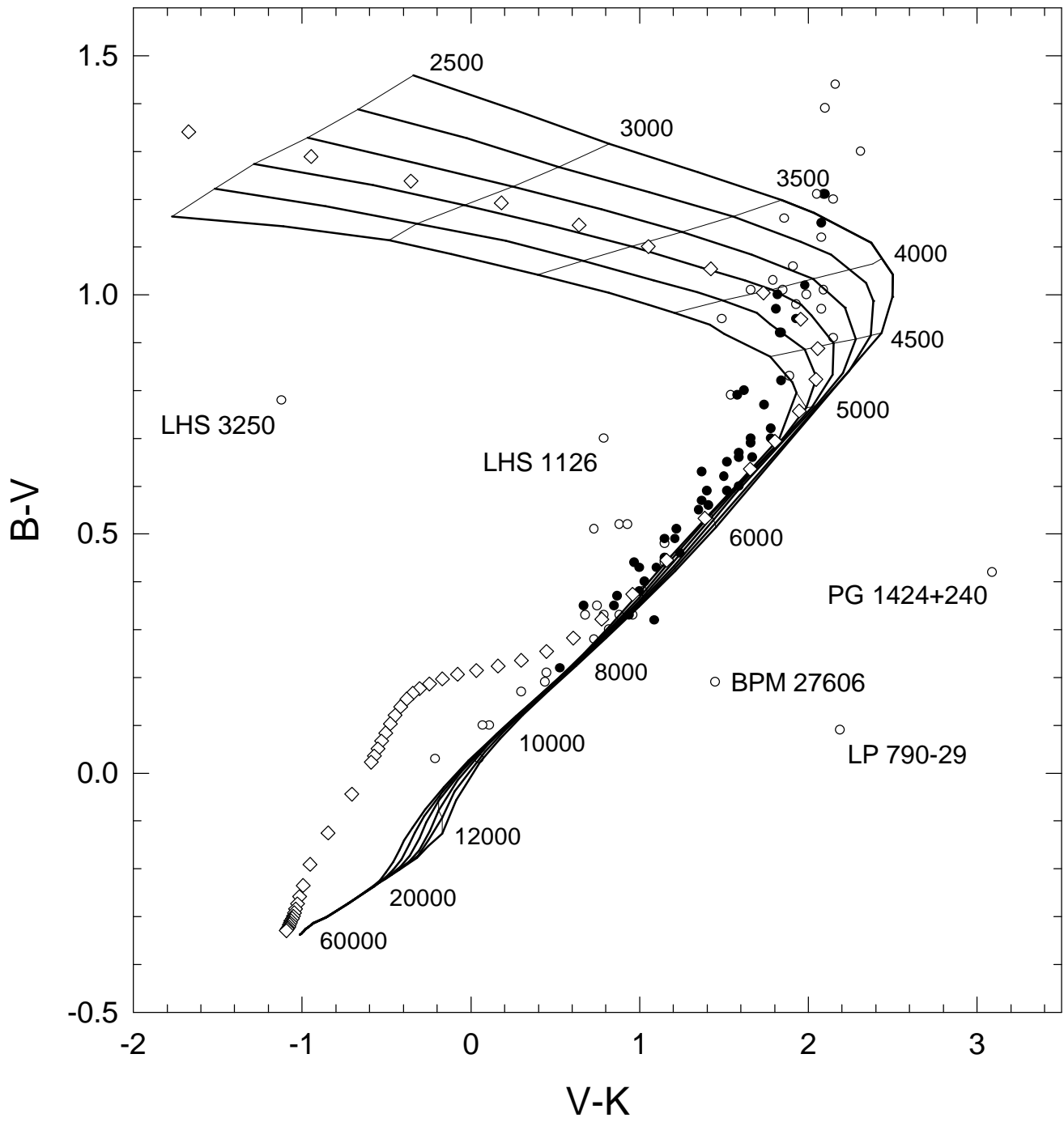
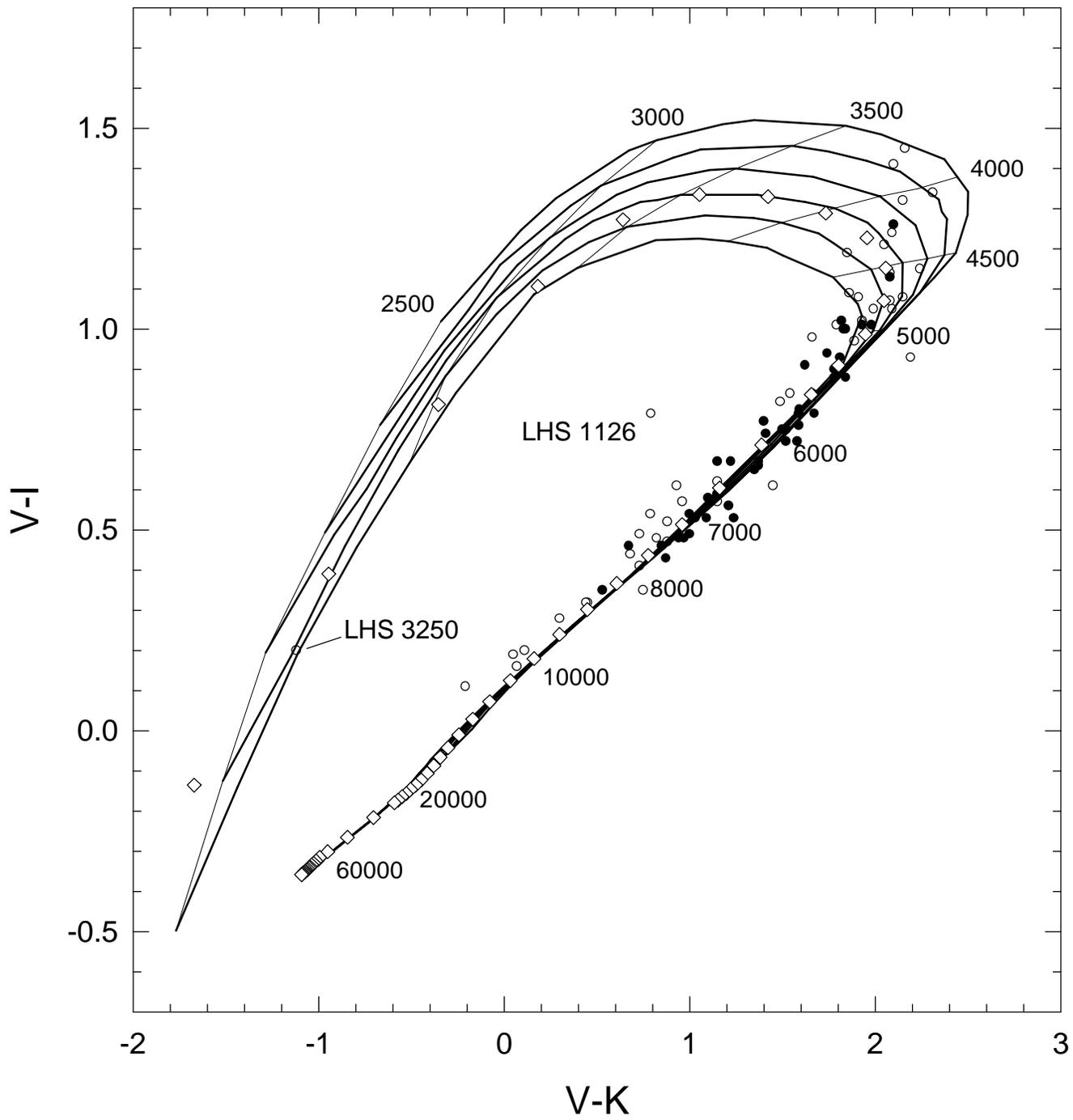
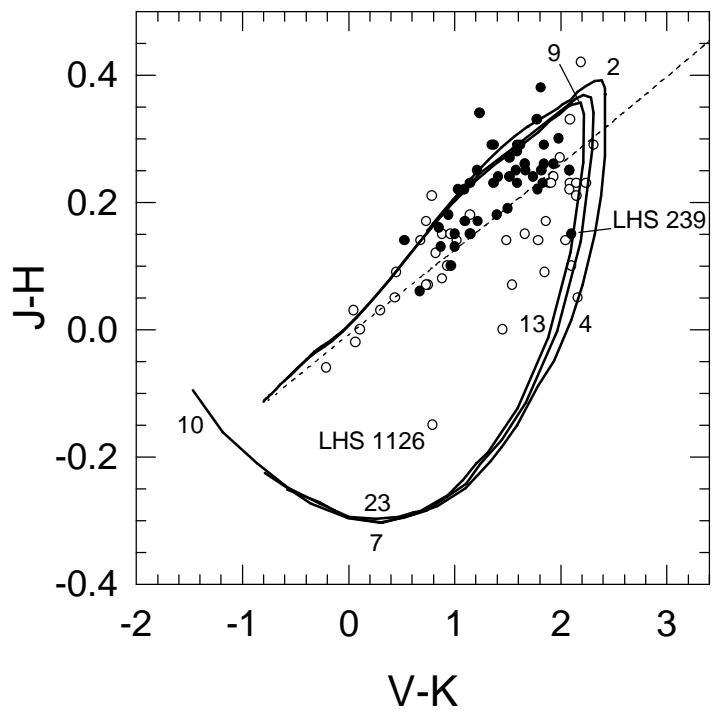
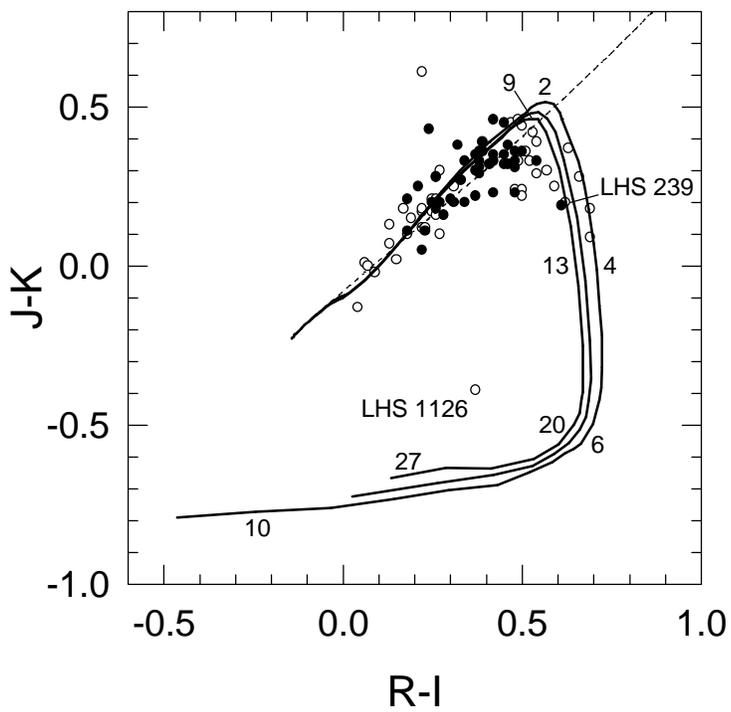
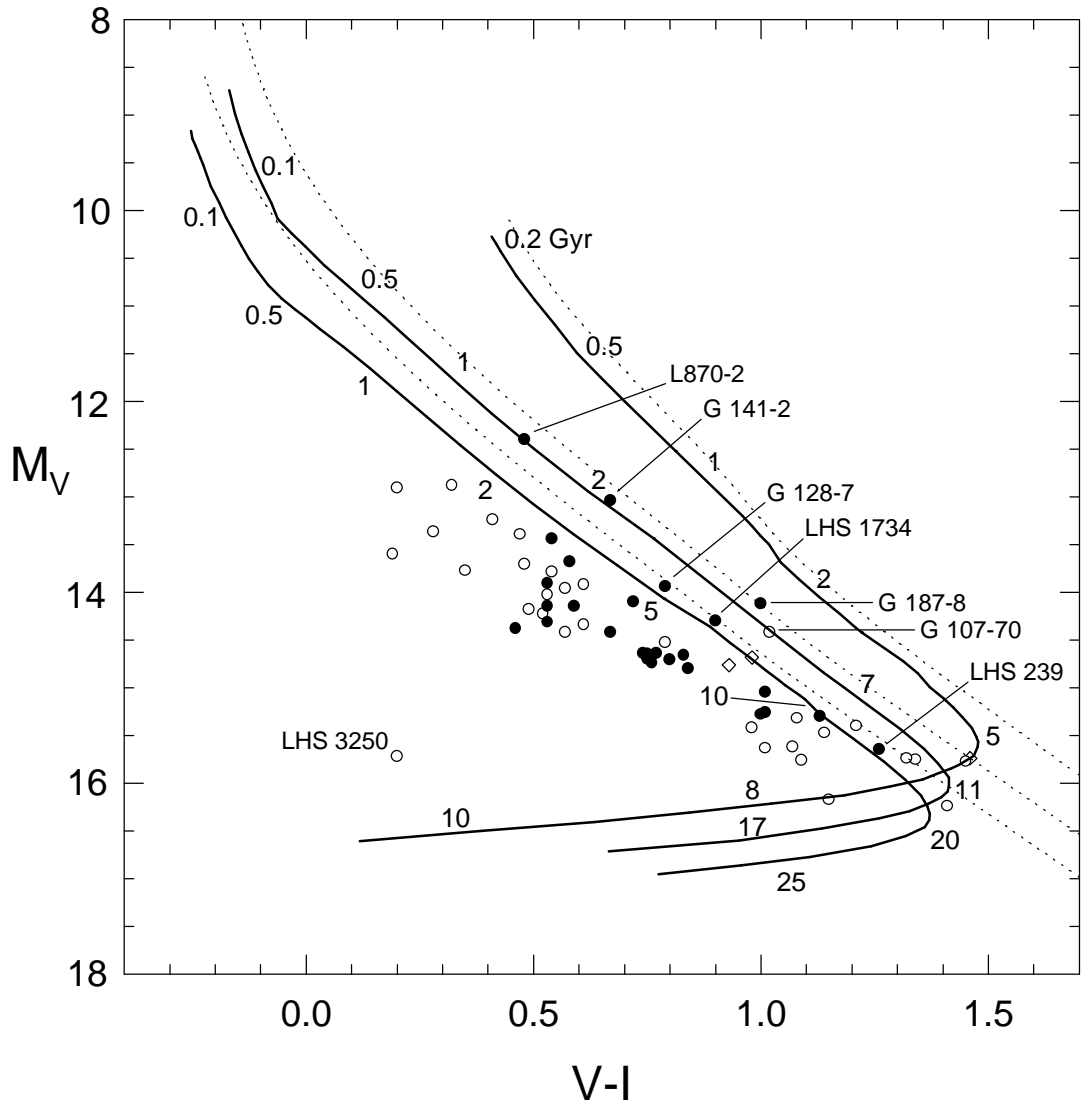
\end{document}